\documentclass[a4paper,fleqn,final]{cas-dc}
\usepackage{xcolor}
\usepackage{caption}
\usepackage[numbers]{natbib}
\usepackage{graphicx}
\usepackage{subcaption}
\usepackage{upgreek}

\usepackage{orcidlink}

\def\tsc#1{\csdef{#1}{\textsc{\lowercase{#1}}\xspace}}
\tsc{WGM}
\tsc{QE}

\begin{document}
\let\WriteBookmarks\relax
\def\floatpagepagefraction{1}
\def\textpagefraction{.001}

\graphicspath{ {Figures/} }

\shorttitle{}    

\shortauthors{}  

\title[mode = title]{Performance of a SuperCDMS HVeV Detector with Sub-eV Energy Resolution and Single Charge-sensitivity}


\author[1]{Kyle Kennard\orcidlink{0009-0003-2394-2666}}
\cormark[1]
\ead{*kylekennard2027@northwestern.edu}

\author[2,3]{Aditi Pradeep\orcidlink{0000-0001-8014-9953}}
\cormark[2]
\ead{**aditi.pradeep@slac.stanford.edu}

\author[4]{Mason Buchanan\orcidlink{0009-0007-1423-5985}}

\author[2,3]{Hope Fu\orcidlink{0009-0009-4626-0231}}

\author[2,3]{Aviv Simchony\orcidlink{0000-0003-4895-3145}}

\author[2,3]{Qihua Wang\orcidlink{0009-0005-8397-7208}}

\author[5]{Emanuele Michielin\orcidlink{0000-0001-5751-8116}}


\author[2,3]{Taylor Aralis\orcidlink{0000-0002-3501-6948}}
\author[4]{Elspeth Cudmore\orcidlink{0000-0002-5824-5145}}
\author[6]{Priscilla Cushman\orcidlink{0000-0002-4950-9807}}
\author[4]{Miriam Diamond \orcidlink{0009-0001-7614-1497}}
\author[1]{Enectali Figueroa-Feliciano\orcidlink{0000-0001-9285-5556}}
\author[7,8]{Caleb Fink\orcidlink{0000-0003-3946-4809}}
\author[4]{Simon Harms\orcidlink{0009-0001-1530-7270}}
\author[9]{Bruce A. Hines\orcidlink{0000-0003-2949-5177}}
\author[4]{Ziqing Hong\orcidlink{0000-0001-8762-4921}}
\author[9,10]{Martin E. Huber\orcidlink{0000-0002-7638-902X}}
\author[11]{Andrew Kubik\orcidlink{0000-0001-9007-422X}}
\author[2,3]{Noah Kurinsky\orcidlink{0000-0002-5872-519X}}
\author[12]{Rupak Mahapatra\orcidlink{0000-0003-0286-1114}}
\author[1]{Valentina Novati\orcidlink{0000-0003-0922-0475}}
\author[13]{Lekhraj Pandey\orcidlink{0000-0003-2199-4289}}
\author[1]{Pratyush K. Patel\orcidlink{0000-0002-2330-2242}}
\author[4]{Weigeng Peng\orcidlink{0009-0002-4062-4347}}
\author[12]{Mark Platt}
\author[4]{Ry Pressman-Cyna\orcidlink{0000-0002-6632-3043}}
\author[14,15]{Wolfgang Rau\orcidlink{0000-0001-9968-5190}}
\author[4]{Runze Ren\orcidlink{0000-0002-4804-4825}}
\author[4]{Tyler Reynolds\orcidlink{0009-0004-0338-7789}}
\author[2,3]{James Ryan\orcidlink{0000-0001-6662-5925}}
\author[16]{Tarek Saab\orcidlink{0000-0003-4463-6011}}
\author[16]{David Sadek\orcidlink{0009-0008-5323-7067}}
\author[1]{Benjamin Schmidt\orcidlink{0000-0001-7118-5936}}
\author[2,3]{Zo$\ddot{e}$ Smith\orcidlink{0000-0003-3886-5352}}
\author[2,3]{Sidney Stevens\orcidlink{0009-0009-8010-0493}}
\author[2,3]{Kelly Stifter\orcidlink{0000-0002-0263-175X}}
\author[11]{Matthew Stukel\orcidlink{0000-0002-9052-7513}}
\author[17]{Julius Viol\orcidlink{0009-0006-9572-4184}}
\author[4]{Yongqi Wang}
\author[18]{Matthew James Wilson\orcidlink{0000-0002-6723-3795}}
\author[19]{Betty Young\orcidlink{0000-0003-4106-9488}}
\author[4]{Stefan Zatschler\orcidlink{0000-0001-8872-5628}}
\author[2,3]{Hazal Zenger}
\author[4]{Ariel Zuñiga-Reyes\orcidlink{0000-0003-4972-5611}}

\affiliation[1]{organization={Department of Physics \& Astronomy, Northwestern University},
            city={Evanston},
            postcode={60208}, 
            state={IL},
            country={USA}}

\affiliation[2]{organization={SLAC National Laboratory},
            addressline={2575 Sand Hill Road}, 
            city={Menlo Park},
            postcode={94025}, 
            state={CA},
            country={USA}}

\affiliation[3]{Kavli Institute for Particle Astrophysics and Cosmology, SLAC National Laboratory, Menlo Park, CA 94025, USA}

\affiliation[4]{organization={Department of Physics, University of Toronto}, 
            city={Toronto},
            postcode={M5S 1A7}, 
            state={ON},
            country={Canada}}

\affiliation[5]{organization={Institute for Astroparticle Physics (IAP), Karlsruhe Institute of Technology (KIT)},
                addressline={76344 Eggenstein-Leopoldshafen},
                Country={Germany}}

\affiliation[6]{organization = {School of Physics \& Astronomy, University of Minnesota}, 
               city = {Minneapolis}, 
               state = {MN},
               postcode = {55455}, 
               country = {USA}}

\affiliation[7]{organization={Institute for Quantum \& Information Sciences, Syracuse University}, 
city={Syracuse}, 
state={NY 13210},
country={USA}}

\affiliation[8]{organization={Department of Physics, Syracuse University}, 
city={Syracuse}, 
state={NY 13210},
country={USA}}

\affiliation[9]{organization={Department of Physics, University of Colorado Denver},
city={Denver}, state={CO 80317}, country={USA}}

\affiliation[10]{organization={Department of Electrical Engineering, University of Colorado Denver}, city={Denver}, state={CO, 80027}, country={USA}}

\affiliation[11]{organization={SNOLAB}, 
            city={Sudbury},
            postcode={P3Y 1N2}, 
            state={ON},
            country={Canada}}

\affiliation[12]{organization={Texas A\&M University},
            addressline={400 Bizzell St}, 
            city={College Station},
            postcode={77840}, 
            state={TX},
            country={USA}}

\affiliation[13]{organization={Department of Physics, University of South Dakota}, city = {Vermillion}, state = {SD}, postcode ={57069}, Country={USA}}

\affiliation[14]{organization={TRIUMF},
            city={Vancouver},
            postcode={V6T 2A3}, 
            state={BC},
            country={Canada}}

\affiliation[15]{Department of Physics, Queen's University, Kingston ON, KL7 3N6, Canada}

\affiliation[16]{organization={Department of Physics, University of Florida}, 
city={Gainesville}, 
state={FL 32611},
country={USA}}

\affiliation[17]{organization={Kirchhoff-Institut für Physik, Universität Heidelberg}, addressline={69117 Heidelberg}, Country={Germany}}

\affiliation[18]{organization={Department of Physics \& Astronomy, University of British Columbia,}, 
            city={Vancouver},
            postcode={V6T 1Z1}, 
            state={BC},
            country={Canada}}

\affiliation[19]{organization={Department of Physics \& Engineering Physics, Santa Clara University},
            city={Santa Clara},
            state={CA},
            postcode={95053}, 
            country={USA}}


\begin{abstract}
We present a detailed characterization of a new generation of athermal-phonon single-charge sensitive Si HVeV detectors, the best of which achieved 589~meV~±~5~meV baseline resolution. Our sub-eV energy resolution enables precise measurements of single-photon events and reveal consistent energy losses of 0.92~eV~±~0.02~eV per charge excitation across two facilities. We demonstrate that the noise for these detectors is well described using a standard Transition Edge Sensor noise model. We also measure a nominal phonon collection efficiency of 45\%~±~3\%~(stat.)~±~6\%~(syst.) in the best performing device, establishing these detectors as the most efficient athermal phonon detectors to date, limited only by intrinsic limitations of quasiparticle generation.    
\end{abstract}



\newcommand{\nk}[1]{\textcolor{red}{(NK: #1)}}
\newcommand{\kk}[1]{\textcolor{blue}{(KK: #1)}}
\newcommand{\ap}[1]{\textcolor{cyan}{(AP: #1)}}
\newcommand{\by}[1]{\textcolor{green}{(BY: #1)}}
\newcommand{\cf}[1]{\textcolor{teal}{(CF: #1)}}



\maketitle

\section{Introduction}
\label{sec:Introduction}

Cryogenic particle detection at the sub-eV scale has progressed substantially in recent years, driven by the needs of low-mass dark matter searches~\cite{Kahn_2022,ZurekReview,Murayama_2023}, precise measurement of nuclear and neutrino interactions at the eV scale~\cite{Augier_2023,nucleus22}. A number of collaborations have now achieved sub-eV resolution~\cite{qrocodile,TESSERACT,UCB25, CRESST_Res}.


The SuperCDMS HVeV program has produced a number of gram-scale detectors with progressively improved phonon-based, single-charge energy detection. Starting with the first demonstration of single-electron resolution~\cite{Romani_2018}, where a detector achieved 10~eV phonon resolution and 0.1~electron-hole pair charge resolution at 100~V, the sensor design and packaging have matured substantially. In Refs~\cite{hong20,ren21}, results with improved sensor geometries and packaging schemes were demonstrated, leading to 3~eV baseline resolution. These new designs enabled detector bias voltages up to 250~V and yielded charge resolution approaching 0.01~electron-hole pairs. Further, a lower bound of 30\% was placed on the phonon collection efficiency. In this paper, we improve on the precision of that measurement using more robust device characterization methods and a better readout scheme than Ref~\cite{ren21} via the use of shielded cables and better separation of TES bias lines from LED lines and improved IR shielding.

These detectors were used to set constraints on electron-recoil dark matter in surface facilities~\cite{Agnese_2018,Amaral_2020} and 
at NEXUS, a shallow-site facility~\cite{albakry25,HVeVR4}. They were used to measure the charge yield for nuclear recoils at a beamline site~\cite{Albakry_2023}, measure the scintillation time constant of a Li$_2$MoO$_4$ crystal~\cite{bartrud}, explore the use of Compton steps for calibrating SuperCDMS Si detectors~\cite{ComptonSteps}, and demonstrate conclusively that scintillation from printed-circuit-boards (PCB) near the detectors was the dominant source of a photon background in earlier searches~\cite{Albakry_2022}. The PCB data led to a redesign of the detector housings, now made of copper, that further improved science results~\cite{HVeVR4}. The results also substantially matured our understanding of charge trapping and impact ionization in cryogenic Si charge transport~\cite{Ponce_2020,Wilson_2024} and were used to improve the accuracy of our detector Monte Carlo through improvements to the underlying simulation framework, G4CMP~\cite{Kelsey_2023} which uses GEANT4~\cite{agostinelli2003, allison2006, ALLISON2016186}. Finally, the calibrations presented here hint at the underlying cause of the discrepancy in previous calibrations of SuperCDMS HVeV detectors performed when operating with and without a high-voltage bias~\cite{Albakry_2023, ComptonSteps}.

In this paper, we report on the device performance of this same detector design with an improved resolution of $\sigma_E$~=~589~meV~±~5~meV compared to our previous result of 2.65~eV~±~0.02~eV~\cite{ren21}. This improved resolution was achieved mostly by lowering the superconducting critical temperature ($T_c$) of the TESs without any other changes to the overall sensor design. In contrast to the older device described in Ref~\cite{ren21}, which had a $T_c$ of $\sim~65$~mK, the new set of detectors have a $T_c$ of $\sim~40$~mK. The lower critical temperature allows for a better athermal phonon resolution. The increased detector sensitivity posed higher challenges on its operation environments, including a stringent requirement on the infrared photon density around the detector. In this paper, we show how upgrades to our device packaging~\cite{albakry25} and readout system have led to successful operation of the new detectors at both the SLAC surface-level test site and at the Cryogenic Underground TEst (CUTE) facility at SNOLAB~\cite{camus18,Camus_2024}. 

A Transition Edge Sensor (TES) based calorimeter reported by {\small TESSERACT} in Ref~\cite{UCB25} demonstrated a  375~meV baseline resolution which was dominated by an unknown source of power noise 
that exceeded the expected sensor noise. This measurement motivated further work to better characterize correlated power noise below trigger threshold. In conjunction with the excess of above-threshold events observed by the dark matter community~\cite{Adari_2022,EXCESS2025}, better tools are needed to study sources of correlated phonon noise, both above and below threshold. Here we present results obtained with an improved SuperCDMS-style HVeV detector that also will be used in subsequent experiments to further study the excess of events above trigger threshold. In this paper, we deploy analysis techniques similar to those used in Ref~\cite{UCB25} to constrain correlated power noise and build a robust model of the detector response. In addition, as the detector sensitivity surpasses the Si band gap scale, the data reveal sub-eV-scale crystal effects including but not limited to various surface effects.

Comprehensive results obtained with low-$T_c$ detectors operated in a dark matter search mode at CUTE (HVeV Run 5) will be presented in an upcoming paper. Here, we show results from two of the low-$T_c$ detectors in experiments designed to validate our noise models and substantially improve our measurements of phonon collection efficiency (first explored in Ref~\cite{ren21}). In Section~\ref{sec:experimental_setup}, we describe detector fabrication and packaging changes made to operate these lower $T_c$ detectors, as well as changes to the readout system that allowed us to operate the detectors at both SLAC and CUTE. In Section~\ref{sec:device_characterization}, we present TES response measurements of the detectors operated at both facilities, and discuss our current understanding of the noise performance of these detectors. In Section~\ref{sec:phonon_calibration}, we illustrate how optical calibration provides valuable data to help model energy loss in HVeV-style detectors with full charge collection. In Section~\ref{sec:discussion}, we conclude by discussing future directions for this style of detector in the context of low-energy phonon backgrounds and other eV-scale science applications.

\section{Experimental setup}
\label{sec:experimental_setup}




The basic unit of our detectors is the W/Al Quasiparticle-trap-assisted Electro-thermal-feedback TES (QET)~\cite{irwin95}. The Al fins trap phonons and direct their energy into the TES which essentially acts as a transducer that converts phonon energy created by an event in the detector substrate to a current that is measured with a Superconducting QUantum Interference Device (SQUID) readout. Hundreds of QETs configured in parallel form a single detector channel. In our design, each QET uses superconducting aluminum thin-films to absorb phonons from the substrate; this energy is then converted into quasiparticles that diffuse in the Al film until they reach W/Al energy ``traps'', where the event energy is concentrated into a voltage-biased tungsten TES operated at a temperature near the middle of its superconducting-to-normal transition. The TES is composed of a $\sim$40~nm-thick W film with a precisely tuned $T_c$ well below 100 mK. More details can be found in Ref~\cite{ren21}. 

The QET and channel design of the devices described in this work were first presented in Ref~\cite{ren21}. The design of these 2-channel detectors is referred to as NFC (short for ``Northwestern-Fermilab C''). We operate the QETs at ground potential, while biasing the (30 nm) Al grid on the opposite side of the substrate at high voltage, establishing an external electric field across the crystal. The devices were fabricated on a 
\(1~\mathrm{cm} \times 1~\mathrm{cm} \times 4~\mathrm{mm}\) Si crystal, oriented with the [100] direction
perpendicular to the 1~\(\mathrm{cm}^2\) face and the side walls oriented perpendicular to [110]. Unlike the devices in Ref~\cite{ren21}, the 1 cm$^2$ faces of the crystal were thermally oxidized to produce a controlled \(40~\mathrm{nm}\)-thick \(\mathrm{SiO}_2\) layer which was included to reduce any potential leakage of charge from the instrumentation into the substrate. The W film was tuned during deposition to a $\sim$40~mK $T_{\mathrm{c}}$ to improve energy resolution.

Results from two test devices fabricated on the same wafer are presented in this paper. One was operated at the underground facility at CUTE, as part of HVeV Run 5. The other was tested in the surface-level facility at SLAC as part of ongoing studies to mitigate vibration-coupled phonon noise and reduce systematic effects in photon-based calibration experiments. The CUTE {\it CryoConcept} dilution refrigerator (DR) is located approximately 2 km underground in the SNOLAB laboratory, near Sudbury in Ontario, Canada~\cite{Camus_2024}. Device testing at SLAC is performed at ground-level in an {\it Oxford ProteoxMX} DR. The mixing chamber stage base temperatures at CUTE and SLAC were around 10 mK (see Table~\ref{tab: basic-params}). 

\begin{figure}[t]
\centering
   \begin{subfigure}[b]{0.45\columnwidth}
        \includegraphics[width=1\textwidth]{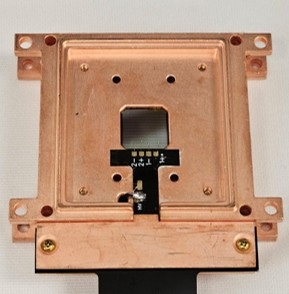}
        \end{subfigure}
\quad
    \begin{subfigure}[b]{0.45\columnwidth}
        \includegraphics[width=1\textwidth]{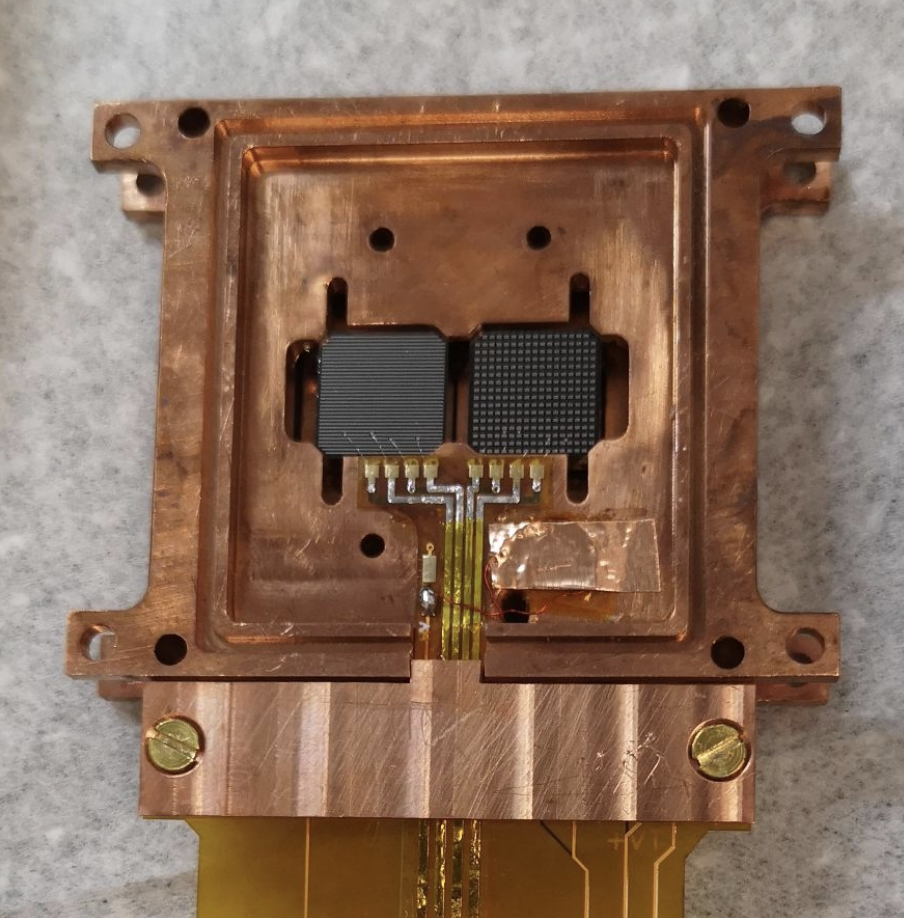}
        \end{subfigure}
    \caption{\textbf{Left} The SLAC mounting scheme for a single device of the NFC design clamped in a copper housing. \textbf{Right} The mounting scheme used at CUTE for two devices (NFC design on the left and NFH on the right) clamped in a similar housing. Both detector housings have an interlocking ridge to create an improved light-tight connection to a lid or other housing (in the case of a stacked payload geometry). The Si detectors shown are 1~cm x 1~cm x 4 mm thick.} 
    \label{fig:det-mount}
    \end{figure}



In both facilities, a copper housing with improved light-tightness was used to hold both the detector and a thin flex-cable for signal readout. At both facilities, the payloads were placed in light-tight cans. Cable feedthroughs at SLAC were potted with silver epoxy, while the feedthroughs at CUTE were epoxied with copper dust filter to further ensure light tightness. 
A two-detector variant of the mounting scheme, first developed for the study in Ref~\cite{albakry25}, was used in the CUTE setup. A single detector mounting variant developed using a slightly modified PCB configuration for convenient coupling to the SQUID readout electronics was used at SLAC. Both housings are shown in Fig.~\ref{fig:det-mount} for comparison.

At CUTE, the detector calibrations were accomplished using in-situ, cryogenic light-emitting diodes (LEDs) with a photon energy of 2.0~eV. The LED was enclosed in a box with a pinhole to focus the light on the center of the face of the HVeV with the Al grid. Each LED package was equipped with infrared filters (\textit{Schott} KG3) to minimize thermal radiation caused by warming the LEDs for operation coupling to the detector. From the mixing chamber plate to the warm stage readout, we used four shielded looms of 25 wires (12 twisted-pairs) each. One loom was used for the TES lines, two for the SQUID lines (one for biasing and the other for the feedback), and one which held the high voltage (HV) bias and LED lines. The stainless steel mesh shielding on each loom was critical for preventing electrical cross-talk between the fast LED control pulses and the other lines. At each thermal stage, feedthrough boards were installed to avoid leakage of infrared photons into the colder stages, with additional application of a mix of Stycast and copper powder on the mixing chamber board. TES signals were carried inside the housings with superconducting wiring to minimize the parasitic resistance. 

At SLAC, the HV and TES signals were carried on a short flex cable with copper traces. This scheme yielded higher parasitic resistance than the CUTE flex cables which were covered in tin-lead solder, providing superconducting connections at detector operating temperatures. The SLAC flex cables were soldered to NbTi twisted pairs that extend to the cold plate stage through an epoxied light-tight feedthrough in an inner mixing chamber shielding can. For the SLAC data runs described here, a stack of six detectors, in separate mounts, was covered by a light-tight Cu lid containing an optical feedthrough fiber for ultraviolet and visible wavelengths for calibration purposes. Using a series of heat sinks and IR filters along the way, the optical fiber ran from the $\sim$10~mK detector stage to a vacuum feedthrough at room temperature, where different LEDs with wavelengths between 280~nm and 810~nm (1.5--4.4~eV) could be connected. In this paper, we focus on the calibration utilizing a 3.4~eV LED in order to better constrain the effects of Si energy band gap losses.

TES signal readout was performed with a single-stage SQUID system specifically designed for HVeV R\&D work, modeled on CDMS designs~\cite{hansen10}. The HVeV system was designed to minimize noise in the TES circuit and calibration systematics from parasitic resistance in the TES-shunt resistor-SQUID input coil loop. This was done in part by putting the SQUIDs and shunt resistors on the $\sim$100 mK cold plate stage, rather than at the $\sim$4 K stage. The SQUID chips used at both SLAC and CUTE (series array chips SA13ax) were fabricated at NIST, and are identical to those used for the SuperCDMS SNOLAB experiment~\cite{huber01}. The CUTE system used 20 m$\Omega$ shunt resistors for the TES biasing circuit, while the SLAC system used 5 m$\Omega$ shunts. Separate 4-wire measurements of the shunt resistors while at operating temperature, which differ slightly from the aforementioned nominal values, are provided in Table~\ref{tab: basic-params}. In general, there is a tradeoff between operating a voltage-biased TES at a lower point in its superconducting transition, to reduce saturation effects (smaller shunt) vs. operating higher in the transition to reduce current noise (larger shunt). The different values of shunt resistors used at the two facilities were chosen to optimize between these two effects. Both facilities used 2~k$\Omega$ SQUID feedback resistors, located on the 4~K stage. SQUID feedback and signal digitizing was accomplished with a SuperCDMS Detector Control and Readout Card (DCRC)~\cite{hansen10}.




At both facilities, substrate bias voltages up to 250~V were generated by an ISEG isolated HV power supply\footnote{EHS 8401p at CUTE, SHR4220 at SLAC}. At SLAC, the fiber-coupled, room temperature LEDs were driven by \textit{Thorlabs} LED drivers, modulated by a \textit{Liquid Instruments Moku:Pro} to produce pulse widths down to the $\upmu$s scale. At CUTE, the cryogenic LEDs were driven using one of the two arbitrary waveform output channels of a NI \textit{Discovery 2 DAQ} device; the second output channel was used to provide a synchronized external trigger. 


\section{Device characterization}
\label{sec:device_characterization}

In this section, we summarize the types of measurements made at each facility to construct a noise and detector response model for subsequent comparison to calibrated performance. First, we measured the complex admittance and thermal response of the TES channels to inform our noise model. We also used TES resistance and bias power data to estimate the phonon energy collection efficiency of the detector, independent of a detailed photon calibration and reconstructed energy spectrum. For validation, we then compare the resulting noise model to measured current noise, from which we produced a predicted final phonon energy resolution and overall phonon collection efficiency. Results of these analyses for detectors operated at both facilities are summarized in Tables~\ref{tab: basic-params} and~\ref{tab: ctii-fit}.


\subsection{QET Characterization}
\label{sub_sec:tes_characterization}
\begin{figure}[t]
\centering
    \includegraphics[width=0.5\textwidth]{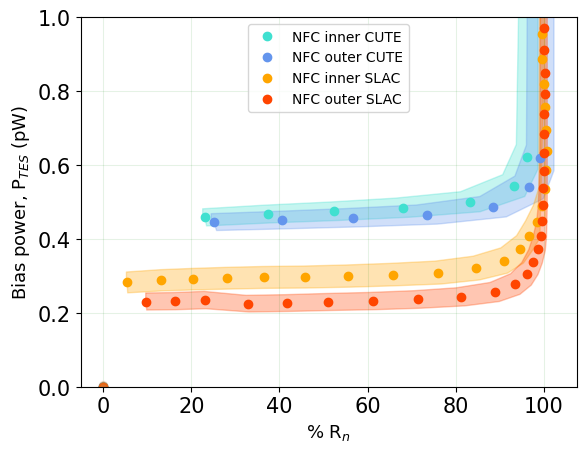}

    \caption{Measured TES bias power $P_0$ = $P_{\text{TES}}$ as a function of TES operating points, defined in terms of relative TES resistance. The shaded regions correspond to $\pm$~1$\sigma$ bands bracketing the data. The error bands are constructed by propagating statistical uncertainties from the measured TES baseline currents and systematic uncertainties dominated by measurements of the shunt resistor in the TES circuit. A fully normal QET channel corresponds to $R_n=100\%$. }
\label{fig:IV_compare}
\end{figure}

In general, to fully characterize the performance of HVeV and similar TES-based detectors, we need to characterize specific TES properties. The TES circuit used in this work is shown in Fig. 3(b) of Ref~\cite{TESBible}. First, we determine the resistance $R_0$ and bias power $P_0 = I_0^2 \cdot R_0$ of the detector channels as a function of operating point within the superconducting transition, where $I_{0}$ denotes the quiescent current through the detector channel at each operating point. We do this by measuring the direct current (DC) from each channel as a function of QET bias voltage ($V_b$). For the devices tested at SLAC and CUTE, the measured TES normal resistance ($R_n$), extracted from resistance profiling, is shown in Table~\ref{tab: basic-params}. The value is consistent between facilities to within measurement uncertainties. However, as shown in Fig.~\ref{fig:IV_compare}, we note that the measured bias power under normal operating conditions is notably larger at CUTE than at SLAC. This result is consistent with our understanding of Joule heating effects in the QETs and the thermal conductance ($G$) of the TESs. The observed difference in bias power could be attributed to environmental factors including but not limited to parasitic power from various sources like noise, vibration, and infrared photons that contribute to the heat load which pushes the TESs from bath temperature\footnote{We assume the crystal to be well-thermalized with the mixing chamber stage of the fridge. Thus, $T_b=T_{\rm mixing\;chamber}$} ($T_b$) to the W film superconducting transition operating point ($T_c$). The value of $G$ under typical operating conditions can be estimated by measuring the TES bias power as a function of bath temperature and applying a thermal balance equation for the TES in equilibrium

\begin{equation}
    P_0 = \frac{G}{nT_c^{n-1}}\left( T_c^n - T_b^n \right) - P_{\rm parasitic},
\end{equation}
where we assume n = 5 for electro-thermal coupling.
 In practice, we cannot directly measure   parasitic power due to vibrations, IR radiation leakage, etc. in the environment. Thus, our thermal power measurements provide only a lower bound on $G$, which can be estimated using: 

\begin{equation}
G =5\cdot T_c^{4}\frac{P_0+P_{\text{parasitic}}}{T_c^5 - T_b^5} \geq {5\cdot T_c^{4}}\frac{P_0}{T_c^5 - T_b^5}.
\label{eq:G}
\end{equation}
\\
\noindent Experimental values for thermal conductance ($G_{\rm chan}$) and $P_{\text{parasitic}}$ are given in Table~\ref{tab: basic-params}. 


As described in Ref~\cite{ren21}, we use raw detector calibration pulses from LED source experiments to directly compute absorbed energy for each event, as well as the overall detector phonon collection efficiency ($\epsilon$), calculated as the ratio of measured-to-expected energy as described in \ref{app:PCE}. The phonon collection efficiency measured at both test sites is summarized in Table~\ref{tab: basic-params}. The calculated efficiencies are consistent with those in Ref~\cite{ren21} for the same design, and with the CUTE measurements we bound the efficiency in the range 39-52\%. 
These measurements confirm the suggestion in Ref~\cite{ren21} that this design is approaching maximum quantum efficiency in phonon readout. These conclusions will be examined in more detail in the discussion.

\begin{table*}[!b]
\caption{Detector parameters for the NFC detectors characterized in this paper. Uncertainties on R$_{p}$, R$_{n}$, L, and P$_{0}$ were dominated by shunt systematics. The corresponding systematic uncertainty was estimated by repeating the analysis many times assuming a shunt resistance sampled from a normal distribution with mean and standard deviation given by the measured value and associated uncertainty from an independent 4-wire measurement. Uncertainties on the thermal conductances reported here are statistical uncertainties. For the SLAC data, the reported thermal conductance is from a single channel and for the CUTE data the reported result is averaged over both channels. The estimation of the uncertainty for the $\epsilon$ is explained in Figure~\ref{fig:efficiency}. The theoretical energy resolutions quoted below is estimated assuming the mean value of the phonon collection efficiency computed at each facility, using the noise data taken with the pulse tube (PT) turned off in both cases.}
\label{tab: basic-params}
  \centering
  \begin{tabular}{llccc}
  \hline
\textbf{Parameter} & \textbf{Description}                   & \textbf{Unit}    & \textbf{SLAC} & \textbf{CUTE} \\ \hline 
$T_b$      & Bath temperature          & mK      & 9-11  & 11 $\pm$ 1  \\
$T_c$      & Measured critical temperature          & mK      & 39  & 49  \\
$R_{\text{shunt}}$      & Shunt resistance          & m$\Omega$      &  4.10 $\pm$ 0.04 & 18.6 $\pm$ 0.9 \\
$R_p$      & Parasitic resistance          & m$\Omega$      &  11.5 & 13.2  \\
\multirow{2}{4em}{$R_n$}     & Outer channel normal resistance             & \multirow{2}{4em}{\centering m$\Omega$} &  390 $\pm$60  & 330 $\pm$ 40   \\
& Inner channel normal resistance  &  & 330 $\pm$ 50 & 320 $\pm$ 30\\
\multirow{2}{4em}{$L$}     & Outer channel inductance             & \multirow{2}{4em}{\centering nH} &  430 $\pm$ 40  &   360 $\pm$ 20   \\
& Inner channel inductance  &  & 450 $\pm$ 50 & 360 $\pm$ 20 \\
\multirow{2}{4em}{$P_0$}      & Outer channel bias power (per channel) at 0.5$\cdot R_n$          & \multirow{2}{4em}{\centering pW }     &  0.23 $\pm$ 0.02    &   0.46 $\pm$ 0.03   \\
& Inner channel bias power (per channel) at 0.5$\cdot R_n$  &     &  0.30 $\pm$ 0.03  & 0.47 $\pm$ 0.03  \\  
$G_{\text{chan}}$         & Thermal conductance (per channel) meas. from bias powers& pW/K    & 22.8 $\pm$ 0.6  & 46.6  $\pm$ 1.3   \\
$G_{\text{TES}}$         & Thermal conductance (per TES) & fW/K    & 45.6 $\pm$ 1.2 &  93 $\pm$ 3  \\
$G_{\text{inferred}}$         & Thermal conductance (per channel) inferred from NEP & pW/K    &  $\sim$83.4  &  $\sim$80.7    \\
$\tau$   & Measured pulse fall time  & $\upmu$s  &  $\sim$140 &   $\sim$200\\
$\sigma_{\text{didv}}$ & Predicted resolution & eV  & $\sim$0.24 & $\sim$0.34 \\
$\epsilon$ & Phonon collection efficiency & \% & 61 $\pm$ 20  & 45 $\pm$ 7 \\
$S_p$ & Noise Equivalent Power (NEP) & aW/$\sqrt{\text{Hz}}$  &   $\sim$3.95 &   $\sim$3.20  \\
$P_{\text{parasitic}}$ & Parasitic power & pW  &  $\lesssim$0.37  & $\lesssim$0.33    \\\hline

\end{tabular}
\end{table*}

\begin{figure*}[t]
\centering
    
    \begin{subfigure}[b]{0.45\textwidth}
        \includegraphics[width=1\textwidth]{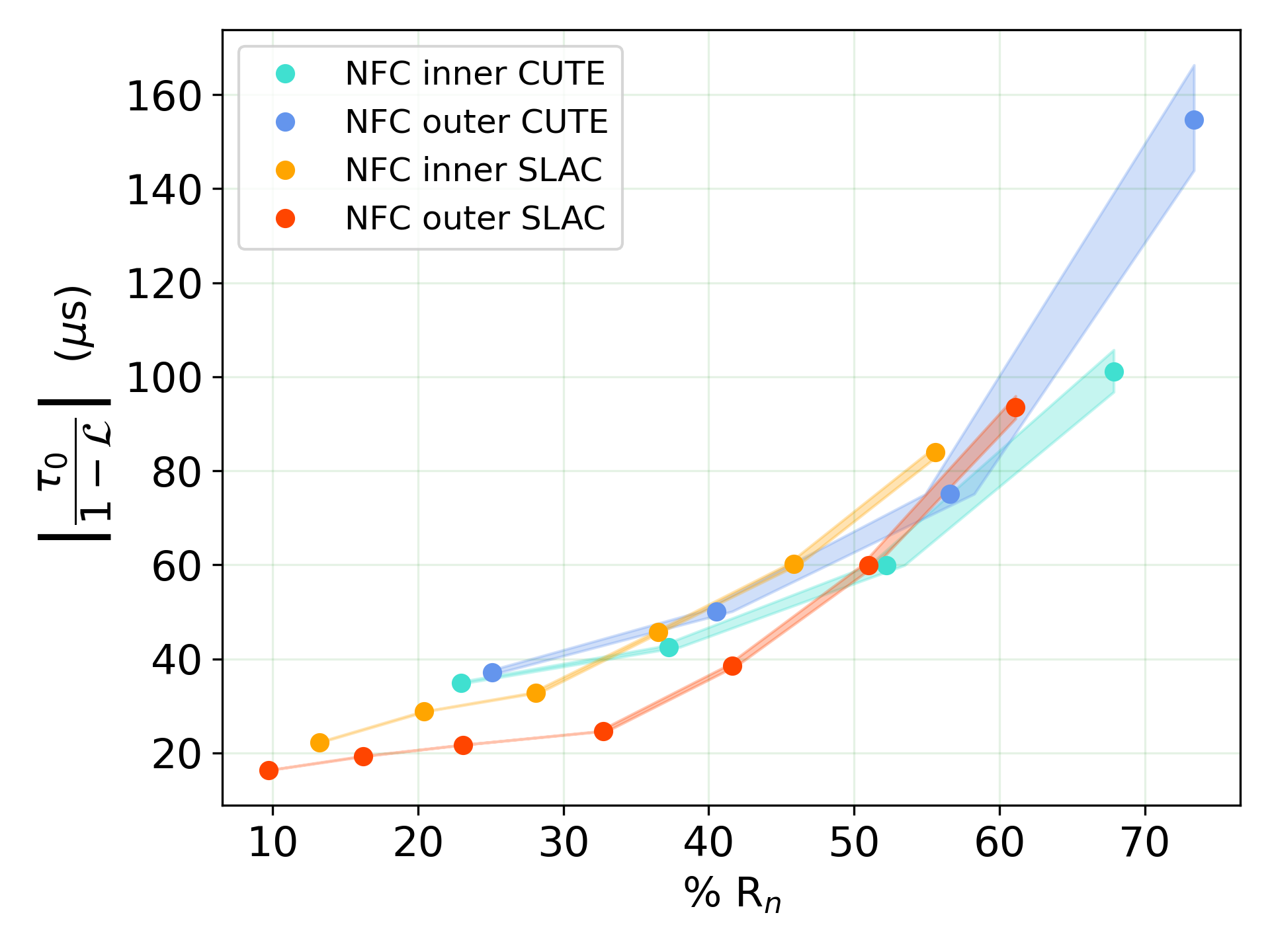}
        
    \end{subfigure}
    \begin{subfigure}[b]{0.45\textwidth}
        \includegraphics[width=1\textwidth]{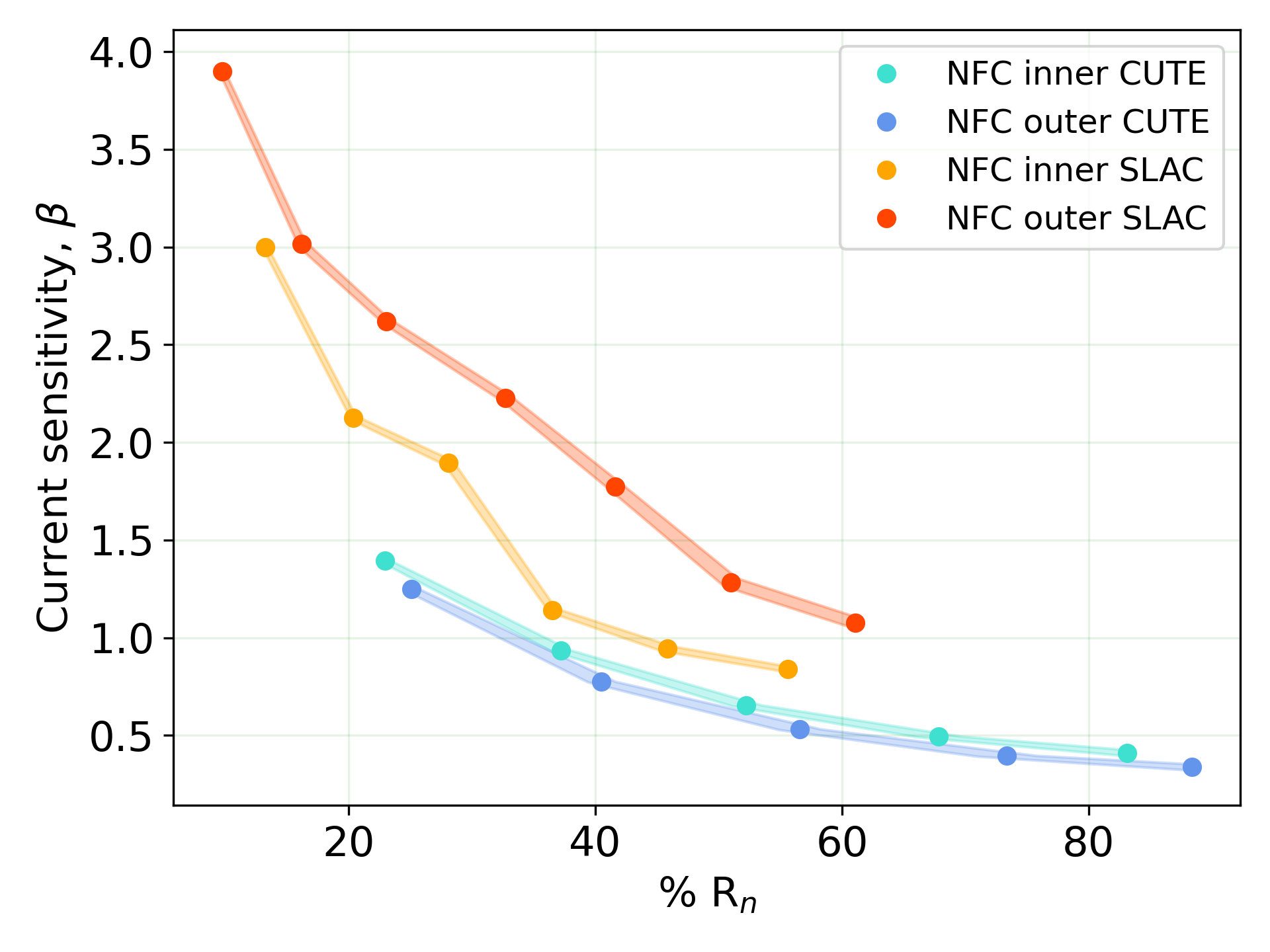}

    \end{subfigure}
    \begin{subfigure}[b]{0.7\textwidth}
        \includegraphics[width=1\textwidth]{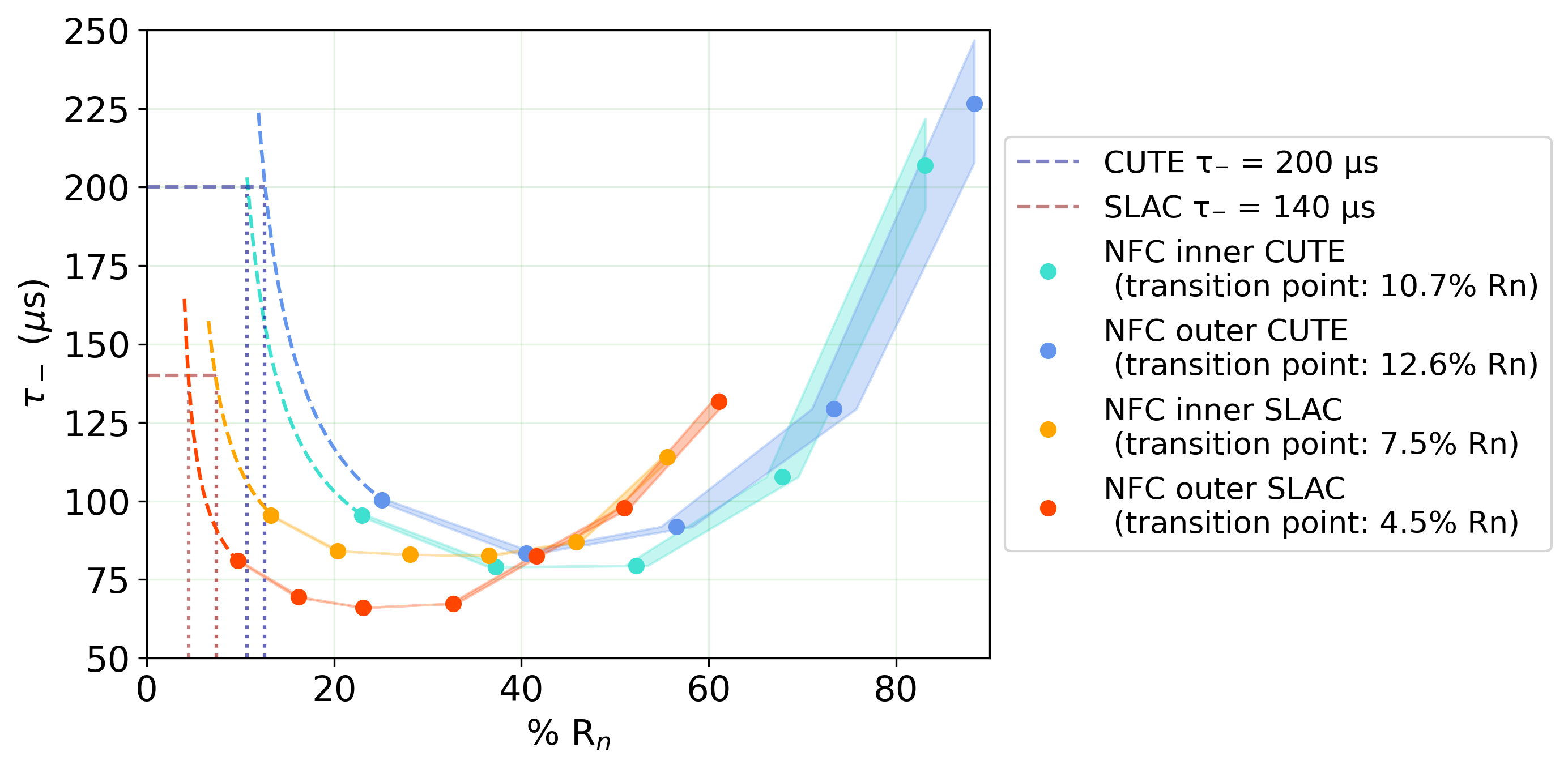}
        
    \end{subfigure}
    \caption{\textbf{Top Left} Constant current TES fall time as a function of the TES bias point (resistance of the TES in steady-state operation here expressed as a percentage of the normal resistance) extracted from fits to the TES complex admittance. \textbf{Top right} Isothermal TES logarithmic current sensitivity $\beta = \frac{d(logR)}{d(logI)}|_T$ as a function of TES bias point. \textbf{Bottom} Fall time of the delta function response in the small signal limit as a function of TES bias point. The points show true measurements made at each facility. The dashed line extrapolates the curve below where data are available from each facility, to compare our predictions with our measured values shown in Table~\ref{tab: basic-params}, confirming that the detectors at both facilities are thermally at lower transition points than expected from the applied bias currents (see text). In each panel, the shaded regions correspond ± 1$\sigma$ bands and are constructed by propagating statistical uncertainties associated with measuring the resistance and systematic uncertainties dominated by measurements of the shunt resistor in the TES bias circuit.} 
\label{fig:Z-results}
\end{figure*}

The remaining elements of the QET response model were determined by fitting the complex admittance of the TES, with the method similar to Refs~\cite{caleb_w_fink_2022_5903353,ren21,UCB25} using the QETpy package~\cite{caleb_w_fink_2022_5903353}. The measurement is performed by applying a small (0.5 $\upmu$A at CUTE and 2.5 $\upmu$A at SLAC) square wave pulse in addition to the quiescent biasing current to the TES bias line and measuring the time dependence of the measured current through the TES, $I_\mathrm{TES}$, assuming a linear response. The complex admittance of the TES, obtained by de-convolving the square wave from the measured $I_\mathrm{TES}$, is modeled as a complex function of the TES resistances, time constants, DC loop gain $\mathcal{L}$, logarithmic current sensitivity $\beta = \frac{\partial(\log{\rm R})}{\partial(\log I)}|_T$, and the inductance in the TES circuit $L$~\cite{Irwin2005,Fink_2020}. The fit results are shown in Fig.~\ref{fig:Z-results}. 

Following the definition in Ref~\cite{Irwin2005}, the fall time shown in the bottom panel of Figure~\ref{fig:Z-results} is
\begin{align}
    \frac{1}{\tau_-} = \frac{1}{2\tau_{el}} + \frac{1}{2\tau_{l}} - \frac{1}{2}\sqrt{\left( \frac{1}{\tau_{el}} - \frac{1}{\tau_{l}}\right)^2 - \frac{4R_0}{L}\frac{\mathcal{L}(2 + \beta)}{\tau_0}},
\end{align}
 where the electronic time constant $\tau_{el} = \frac{L}{R_l + R_0(1+\beta)}$, the electro-thermal feedback time constant $\tau_{l} = \frac{\tau_0}{1-\mathcal{L}}$ and the natural TES time constant $\tau_0 = \frac{C}{G}$ contribute. 
The figure shows good agreement between the predicted fall times ($\tau_-$) and the fall time measured with an exponential fit to an average pulse at the two facilities ($\tau$). We also find a decreasing current sensitivity with increasing TES temperature, as seen in past devices~\cite{Fink_2020}. The inferred phonon pulse fall times, shown in the top left, are shorter than the measured template fall times at the respective bias points the detectors are set to. One possibility is that the discrepancy may be due to the pulse data being taken lower in transition than what the detectors are set to, as fall time rises for pulses close to the point where the electronic time constant and TES response time become comparable. The phonon dynamics are expected to occur on the tens of microseconds scale, as described in Ref~\cite{ren21}, and are not expected to affect TES response time.

\subsection{Noise Modeling}
\label{sub_sec:noise_performance}

The studies shown in Sec. \ref{sub_sec:tes_characterization} allow us to move to fitting a noise model to data with a minimum number of free parameters to allow for more meaningful interpretation of device performance. The QET noise is modeled as comprising three components that combine in quadrature: (a) readout noise from the SQUIDs with a characteristic $1/f$ dependence at low frequency, (b) white power noise originating from thermal fluctuations across the interface between the TES and the surrounding thermal bath, and (c) Johnson noise from the TES and from other passive elements in the circuit. 

Noise in the inductively coupled SQUID amplifier sets the noise floor for the current readout, and constitutes a fixed current noise in excess of the current noise present in the TES loop itself. We can  constrain the model of the SQUID noise by considering the case where the TES is in its fully resistive state.
Similarly, Johnson noise contributions from the passive elements are modeled with the TES in its superconducting state. 
When the TES is in transition, its small but finite resistance ($<$ 0.5~$\Omega$) introduces Johnson noise and non-zero thermal fluctuation noise that is dependent on the electro-thermal properties of the TES like $\tau_0, \,\beta, \,\mathcal{L}$ and inductance (L) of the TES circuit which are extracted from the measured TES complex admittance data. Example results from this measurement are shown in Fig.~\ref{fig:Z-results}.

The noise analysis is performed in the frequency domain by computing amplitude spectral densities (A/$\sqrt{\rm Hz}$) from the averaged Discrete Fourier Transform (DFT) of events consistent with noise that are extracted from data at a fixed rate of a few Hz. 
To convert current noise into noise equivalent power (W/$\sqrt{\rm Hz}$), we use the current-to-power transfer function, $|{\delta I}/{\delta P}|$ derived from modeling the complex admittance of the TES circuit~\cite{Fink_2020}.

\begin{figure*}[t]
    \centering
    \includegraphics[width=0.95\textwidth]{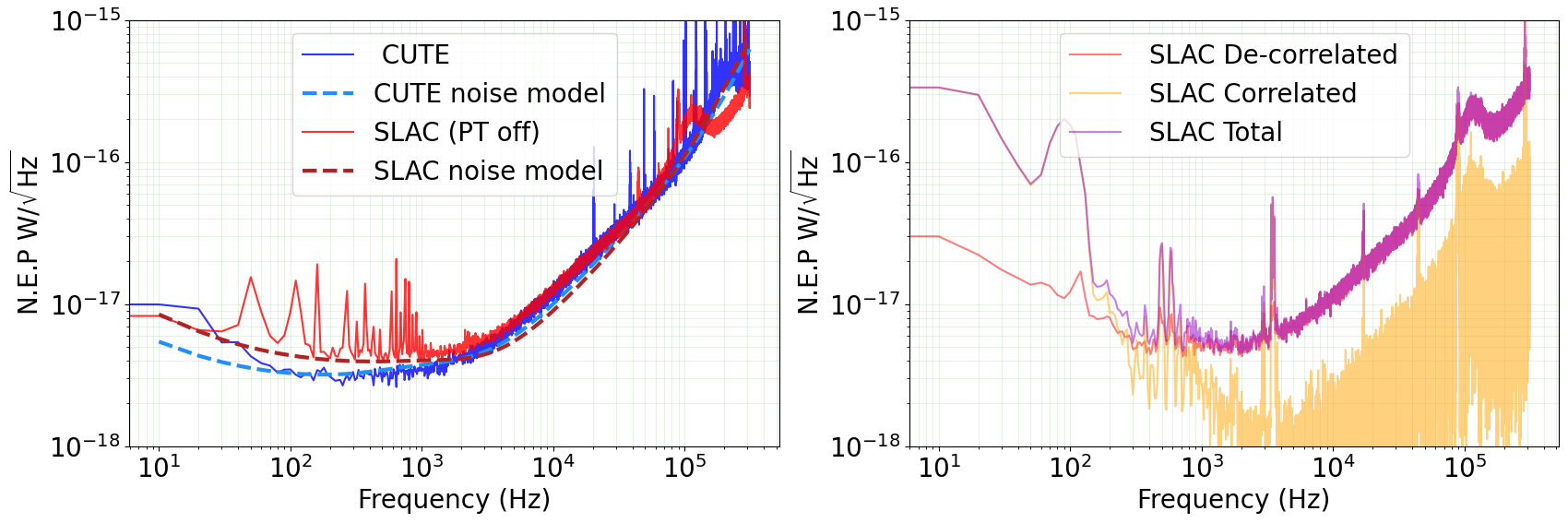}
    \caption{\textbf{Left}: A comparison of the measured Noise Equivalent Power for a single QET channel at the two test facilities (SLAC and CUTE). The left panel compares the lowest-noise spectrum measured at each of the two facilities along with their corresponding best fit noise models, represented by a quadrature sum of the four noise components described in Section~\ref{sub_sec:noise_performance}. The red spectrum was computed using SLAC data acquired when the pulse tube (PT) of the cryostat was turned off, to minimize vibrational noise coupling into the detector. Contributions from environmental noise at 60~Hz (and its harmonics), as well as digital communication noise in the kHz regime are evident. Work to mitigate these noise sources is ongoing. \textbf{Right}: Total (purple), decorrelated (red) and correlated (yellow) noise spectra obtained at SLAC with the PT cooler turned on. The dominant correlated noise contribution seen below $\sim$300~Hz is due to the PT. Mechanical resonances associated with the cryostat are also evident in the spectra near 3~kHz. They do not appear in the PT off spectrum shown in the left panel.}
    \label{fig:power_noise}
\end{figure*}

We compare the measured noise power with our noise model to benchmark the noise performance of the device as shown in Fig.~\ref{fig:power_noise}. In the model, only the thermal conductance ($G$) is treated as a free parameter. The predicted white power noise is then:
\begin{align}\label{eq:TFN}
    S_{p}^{1/2} &= \sqrt{4k_BT_cG} \\
    &\approx 8.6\cdot10^{-18}\mathrm{\frac{W}{\sqrt{Hz}}}\left[\frac{T_c}{45~\mathrm{mK}}\cdot\frac{G}{30~\mathrm{pW/K}}\right]^{1/2},
\end{align}
where $G = G_{\text{inferred}}$ from Table~\ref{tab: basic-params}. The lower limit of the white power noise is set using the thermal conductance computed from the bias power, with results summarized in Table~\ref{tab: basic-params}. A key finding is that the measured power noise (N.E.P) varies by $\sim$20\% between facilities in the region dominated by thermal fluctuation noise. This difference is explained by the three-pole nature of the complex admittance fit which points to an additional heat capacity caused by two or more thermal links between the TES, absorber and the bath~\cite{maasilta_2012}. The measured noise for each facility is described well by the TES noise model for values of $G_{\text{inferred}}$ reported in Table~\ref{tab: basic-params}. The $G_{\text{inferred}}$ from both SLAC and CUTE are consistent at the 3\% level. This suggests that power noise modeling provides a viable unbiased method to estimate the thermal conductance G, comparing to the lower limit calculated using Equation~\ref{eq:G}. We have also used the inferred G value from this noise model to put an upper bound on parasitic power for both facilities, with results shown in Table~\ref{tab: basic-params}.

Differences in the vibrational environments at the two test facilities produced observable differences in the noise performance of the devices. These discrepancies primarily arise due to correlated noise between channels caused by vibrational couplings into the detector from the pulse tube coolers in the cryostats. The effect is clear when comparing the PT-on noise measurements in the right panel of Fig.~\ref{fig:power_noise} to the PT-off measurement shown in the left panel.
In CUTE, the detector payload is installed in a cryostat that is mechanically decoupled from the vibrationally active components of the facility and from environmental sources of vibration~\cite{Camus_2024}. In contrast, the SLAC payload was mounted directly onto a cryogenic vibration isolation platform\footnote{For vibration isolation at SLAC, we utilized a JPE CVIP3 cryogenic isolation stage}, that is suspended from the mixing chamber plate in the cryostat. The commercial isolation platform is only effective in suppressing vibration noise above $\sim$100 Hz. To mitigate the low frequency vibrational couplings that can dominate the SLAC noise, we compute the Cross Spectral Density (CSD) matrix of the two detector channels. At every frequency, this method separates the uncorrelated noise of each channel (captured in the diagonal elements) from the cross-correlated noise between channels. We then subtract the inverse CSD matrix from the total measured noise, to compute a more robust correlated noise estimator for each channel (see \ref{app:appendixD} for more information). 

We note that this procedure is subject to large uncertainties in cases where the decorrelated noise is more than an order of magnitude lower than the total noise, due to the uncertainties in the measurements of the noise spectra. Therefore, as shown in left panel of Fig.~\ref{fig:power_noise} for the SLAC run analysis, we compared the lowest measured noise powers when the PT was turned off, to confirm that we approach the noise level measured at CUTE when we remove the pulse tube induced vibration in the system. Residual noise in the PT-off data in this case could be from electrical pickup present during the period that the PT-off noise was collected.

\subsection{Predicted Energy Resolution}


One benefit of the characterization procedure described above is that it provides a direct prediction of the energy resolution $\sigma$, given a noise power spectrum representative of the true detector noise, $N^2(f)$, and the expected detector pulse response in the small signal limit, $s(f)$,
\begin{equation}
    \label{eq:OF_resolution}
    \sigma^2 = \left [ 4 \int_0^\infty \frac{|s(f)|^2}{N^2(f)}\right].
\end{equation}
Since thermal fluctuation noise dominates the total noise for each TES, Eq.~\ref{eq:OF_resolution} can be expressed in terms of experimental parameters as:
\begin{equation}
    \label{eq:theory_res}
    \sigma \approx \frac{T_c^3}{\epsilon}\sqrt{2n\Sigma\frac{v_{\text{TES}}}{\zeta_{\text{TES}}}k_b(\tau_{ph} + \tau_-)} ,
\end{equation}

\noindent where $T_c$ is the tungsten TES film critical temperature, $\epsilon$ is the phonon collection efficiency, $\Sigma$ is the electro-thermal coupling constant, $v_{\text{TES}}$ is the TES volume, $\zeta_{\text{TES}}$ is the fractional volume of the TES film separate from the Al/W trapping regions, $\tau_{ph}$ is the phonon collection time, $\tau_-$ is the TES response time, and $n = 5$ is the thermal conduction power-law exponent used in the model.


Experimental values of the fit parameters obtained from the best power noise spectrum from each facility are shown in Table~\ref{tab: basic-params}. 
Much of the improvement in resolution shown in this work compared to earlier experiments~\cite{ren21} can be attributed to a reduction in W film $T_c$ from $\approx$~65 mK to 40~mK. Furthermore, this work produced more stringent constraints on the phonon collection efficiency than earlier work, by reducing systematic uncertainties associated with the value of the shunt resistor  at cryogenic temperatures by performing independent 4-wire measurements of the shunt resistance as a function of temperature. Combined, these improvements enabled us to constrain the energy resolution expected from this detector. The predicted resolutions from the two facilities computed with the mean phonon collection efficiency using noise data taken with the pulse tube off are shown in Table~\ref{tab: basic-params}.

\section{Optical Calibration Results}
\label{sec:phonon_calibration}

In the previous section, we constructed a model of QET performance largely informed by independent measurements of TES properties, with final detector energy resolutions predicted from rough calibrations, as discussed in \ref{app:PCE}. 
The intent was to ensure that any optimal filter-based energy resolution found through our usual data-driven calibration procedure with events induced by optical photons will be robust.

\subsection{Energy Resolution}
As discussed in Section~\ref{sec:experimental_setup}, the detectors at both SLAC and CUTE were calibrated using optical LED photons. The light was aimed at the polished backside of the detector that was not instrumented with QETs. In the CUTE setup, the measured LED photon energy was 2.0468 $\pm$ 0.0002 eV (605 nm). The measurement was taken with a spectrum analyzer when the LED was immersed in liquid helium. During operation, a 1~$\upmu$s voltage pulse from a 500 kHz sinusoidal was sent to the LED to emit a small number of photons every 100 ms. The number of photons emitted during each pulse followed a Poisson distribution with mean $\lambda$ that scaled with the amplitude of the voltage signal sent to the LED. An external trigger coincident with the rising edge of the voltage pulse was used to identify detector phonon signals induced by the LED photons.

\begin{figure*}[t]
\centering
    \begin{subfigure}[b]
    {0.48\textwidth}
        \includegraphics[width=1\textwidth]{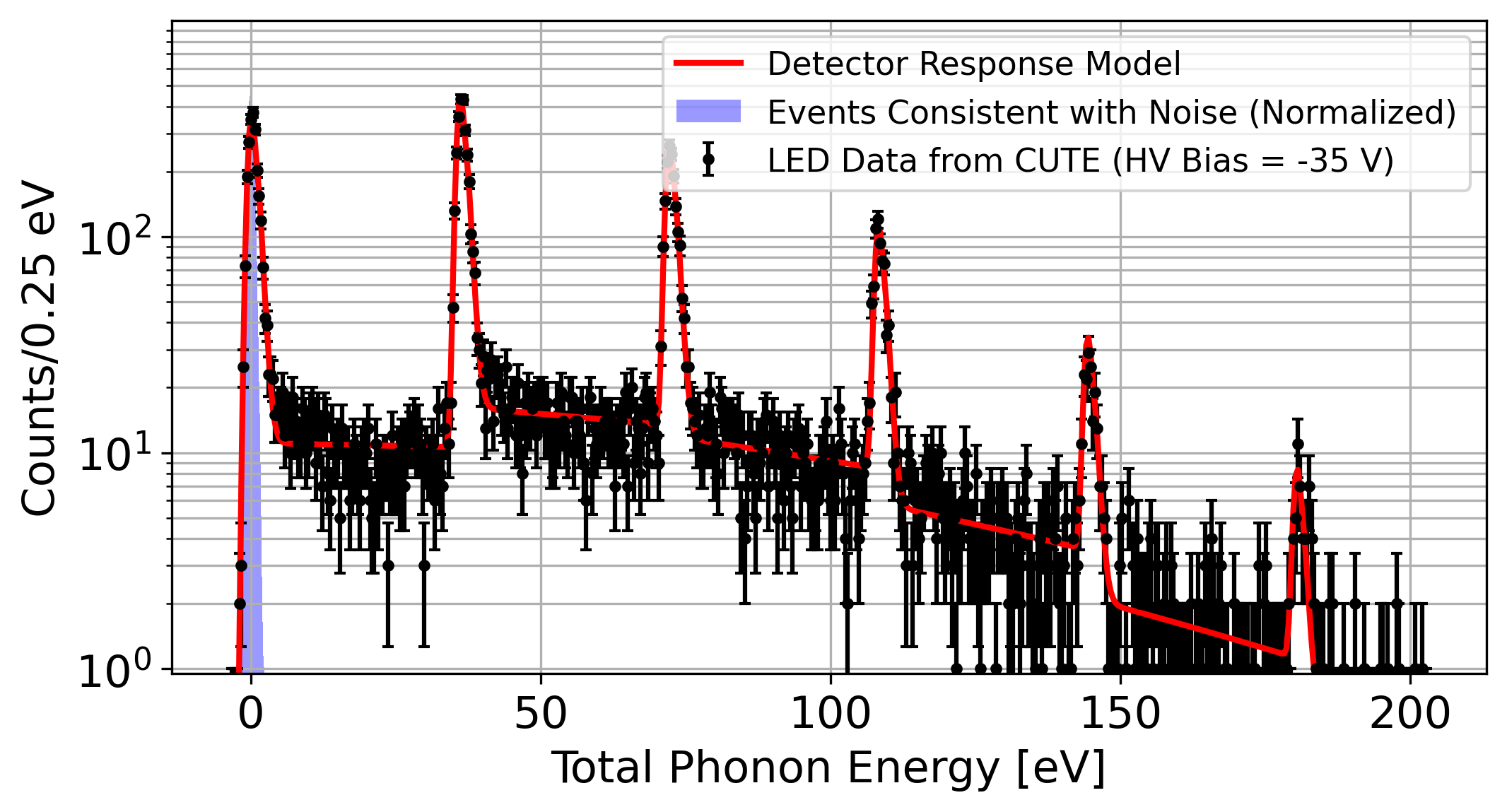}
        
    \end{subfigure}
    \begin{subfigure}[b]{0.48\textwidth}
        \includegraphics[width=1\textwidth]{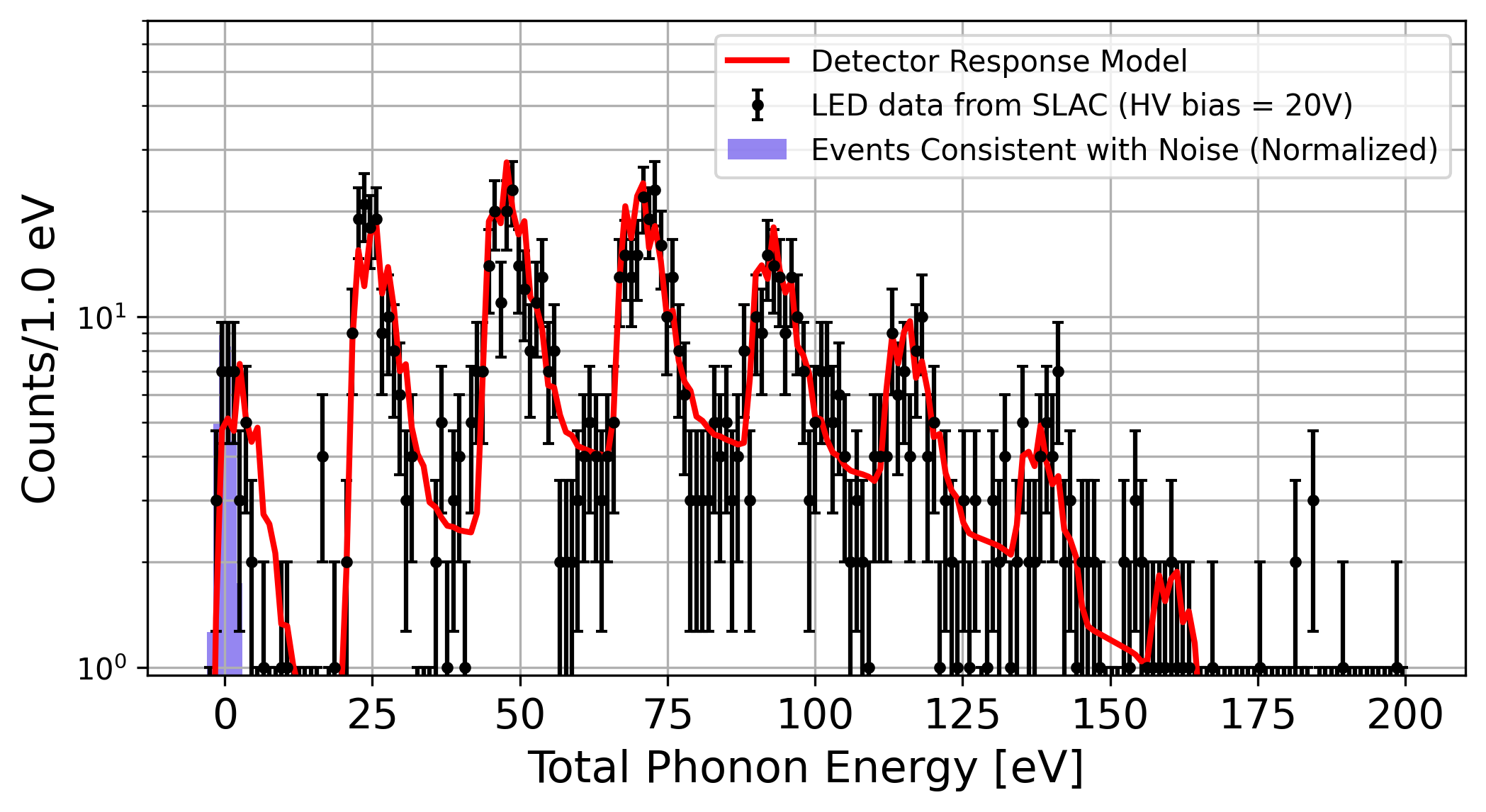}
        
    \end{subfigure}

\caption{\textbf{Left:} Cross-talk-subtracted LED energy spectrum (black) from a data set taken at CUTE with 2.0~eV photons. Error bars for the y-axis represent the standard deviation of a Poisson distribution with the mean number of counts equal to that of a given bin. The fit shown in red corresponds to the detector response model described in this work. The energy distribution for events consistent with noise is shaded in blue and has a characteristic width consistent with the resolution returned by the detector response model fit. \textbf{Right:} The energy spectrum of LED events at SLAC with 3.4 eV photons. Note that the jagged features on the peaks are due to a combination of surface trapping and recombination effects, as well as the inequality between charge production and photon number. They are visible here (and not the left panel) because the collected photon energy is sufficiently large compared to the resolution, as discussed in Sec. \ref{photon_energy_loss}.}
\label{fig:led_cal}
\end{figure*}

The detector energy scale and phonon resolution were found using an optimal filter (OF) analysis which is discussed in detail in  \ref{app:Trigger}. 
For data taken at the CUTE facility, a correction to the pulse amplitude measured with the optimum filter at the LED incident time (A$_{\text{OF}0}$) was made to account for cross-talk (see \ref{app:crosstalk}). To reduce the degeneracy of the parameter space in the detector response model fit to the calibration spectra, described in Ref~\cite{Wilson_2024}, four data sets with varying LED voltage amplitudes were fit simultaneously. Parameters dependent on LED operation and behavior, in this case the Poissonian mean photon number $\lambda$ per photon pulse and the overall normalization of the detector response model, were allowed to vary across data sets. Other model parameters, $e.g$, energy calibration coefficients, phonon signal resolution, and charge mobility parameters -- including the probability for surface trapping, charge trapping, and impact ionization -- were shared across data sets. These parameters are listed in Table~\ref{tab: ctii-fit}. We used a Markov-chain Monte Carlo (MCMC) to ensure convergence of this fit with the large parameter space. 

At SLAC, the detector energy calibration and measurement of the phonon signal energy resolution were made using 365~nm (3.40 eV) LED photons following a procedure similar to that at CUTE. The SLAC data did not require a cross-talk correction because the LED light was coupled to the detector payload via optical fiber. The SLAC fit parameters were less constrained than those for CUTE, due to lower statistics, and data with a single (Poisson) $\lambda$ value rather than multiple values, as was the case at CUTE. 

\setlength{\extrarowheight}{2pt}
\begin{table*}[!b]
\caption{Fitted parameters from the full model for HVeV detector calibration used in this paper. }
\label{tab: ctii-fit}
  \centering
  \begin{tabular}{lccrr}
  \hline
\textbf{Parameter} & \textbf{Description}             & \textbf{Unit}    & \textbf{SLAC} & \textbf{CUTE} \\ \hline 
$\sigma_m$      & Resolution          & meV     & $900_{-2}^{+22}$ & $589 \pm 5$  \\ [2pt] 
$P_{\text{CT}}$   & Charge trapping probability  & \%  &  $10_{-3}^{+2}$ & $27.2 \pm 0.2$ \\ [2pt] 
$P_{\text{II}}$ &Impact ionization probability &\% & $7_{-2}^{+3}$ & $1.1 \pm 0.4$  \\ [2pt] 
$P_{\text{ST}}$ & Surface trapping probability & \%  & $29 _{-2}^{+1}$ & $21.9 \pm 0.6$ \\ [2pt] 
$E_{\text{ph}}$ & Collected photon energy & eV & $2.6_{-0.2}^{+0.1}$ & $1.13 \pm 0.02$ \\ [2pt] 
$c1$ & Linear calibration factor & eV/$\upmu \text{A}$  &  $290_{-4}^{+2}$ &  $339.3 \pm 0.2$ \\ [2pt] 
$c2$ & Second-order calibration factor & eV/$\upmu \text{A}^2$   &   $27 _{-4}^{+5}$ &  $55.9 \pm 0.4$\\ [2pt] \hline
$E_{\gamma}$ & Photon energy & eV  &  3.40 & $2.0468 \pm 0.0002$ \\
$E_{\text{loss}}$ & Lost photon energy & eV  &  $0.8_{-0.1}^{+0.2}$ & $0.92 \pm 0.02$ \\ [2pt] \hline

\end{tabular}
\end{table*}

Separate energy calibration histograms from CUTE and SLAC data sets are shown in Fig.~\ref{fig:led_cal}, along with fits to the histograms. Energy calibration values for these data sets are presented in Table~\ref{tab: ctii-fit}, where we show the mean values and standard deviation of the posterior sample from each fit.

\subsection{Constraints on Charge Mobility and Collected Photon Energy}
\label{photon_energy_loss}
Histogram peaks in the precision detector calibration data presented in Fig.~\ref{fig:led_cal} 
were individually fit to a simple model:
\begin{equation}\label{eq:phononEnergy}
    E_{n} = n_e\cdot V + n_{\gamma}\left[E_{\gamma} - E_{\text{loss}}\right],
\end{equation}
where $n_e$ is the number of electron-hole pairs produced by a single LED event, $n_{\gamma}$ is the number of photons in the LED pulse that were absorbed by the detector for that event, $E_{\gamma}$ is the incident photon energy, $E_{\text{loss}}$ is the fraction of initial photon energy not recovered during the event, and $V$ is the HVeV detector substrate bias. Data sets such as those shown in Fig.~\ref{fig:led_cal} allow us to explore multiple limits of the simple fit equation (Eq.~\ref{eq:phononEnergy}), and determine quantitative relationships between $n_e$ and $n_{\gamma}$, and between $E_{\text{loss}}$ and $E_{\gamma}$. The basic model has been used successfully to estimate specific charge-mobility properties in HVeV devices. Its application is described in more detail in Ref~\cite{Wilson_2024}.

The quantized n-electron-hole-pair peaks fitted in Fig.~\ref{fig:led_cal} have high-energy tails that provide important information about the incident photon distribution. The tails are also useful for constraining the number of electron-hole pairs generated per photon event. In addition, by using the MCMC analysis described earlier, we can estimate charge-trapping (bulk and surface) and impact ionization effects as a function of detector bias and other operating conditions. Differences in charge trapping, impact ionization, and surface trapping probabilities between CUTE and SLAC data in Fig.~\ref{fig:led_cal}
are likely due to the different voltage biases under which the data were taken. We note also that the reconstructed photon energy for both the CUTE and SLAC datasets is less than the incoming photon energy. 
Notably, in the SLAC data, there is some probability of one photon resulting in either one or two electron-hole pairs, producing a more complex event distribution above each electron-hole pair peak.
 
\begin{figure}[t]
\centering
    \includegraphics[width=0.47\textwidth]{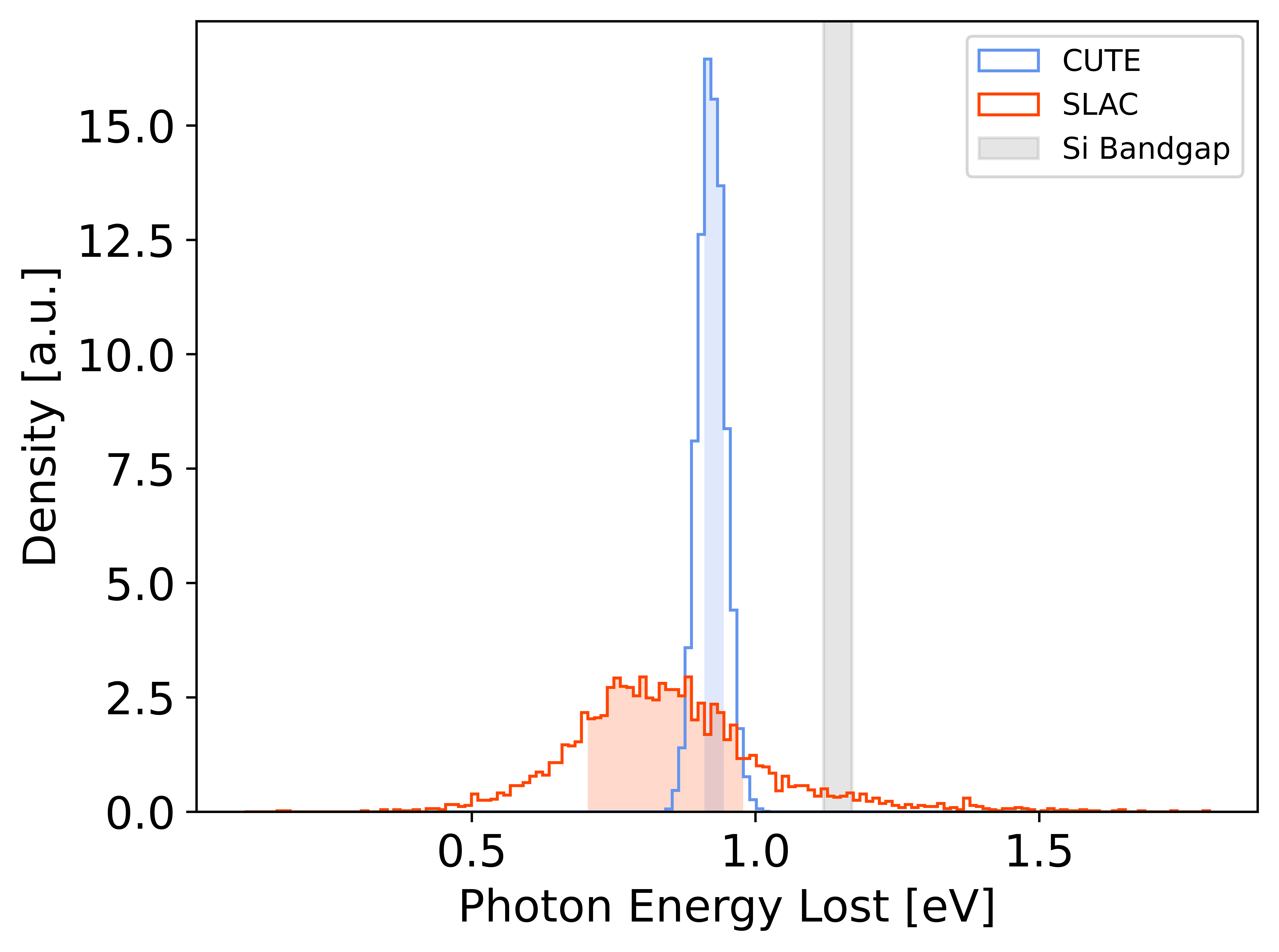}

    \caption{Samples from the posterior distributions of event energy loss per photon using LED data from SLAC (red) and CUTE (blue). The corresponding shaded regions represent ±1 standard deviation from the mean. The posterior distribution from CUTE results from the simultaneous fit to four data sets whereas the distribution from SLAC results from a fit to a single data set. The gray band shows the band-gap of silicon~\cite{ramanathan2020ionization}. The narrowness of the distribution from CUTE compared to SLAC is a result of using four data sets to constrain the parameter space. The distributions show consistent results between data taken at CUTE and SLAC.}
\label{fig:e_loss}
\end{figure}

The best-fit models reveal that statistically-significant non-zero $E_{\text{loss}}$ terms are required to adequately model the data.
The resulting best-fit model values for the data presented here are shown in Table \ref{tab: ctii-fit}. The posterior fit distributions for the $E_{\text{loss}}$ term are shown in Fig.~\ref{fig:e_loss}. The fit results suggest that, on average, a large fraction of the bandgap energy is not recovered. Future work is needed to more robustly characterize this effect as a function of bias voltage and LED wavelength. We will discuss in the next section potential implications for this $E_\text{loss}$, including corroborating the previously seen energy-scale discrepancies when comparing HVeV data sets with 0~V vs. high-voltage bias.

\section{Discussion}
\label{sec:discussion}

In this paper, we have shown the substantial improvement made in HVeV detector performance and modeling benefiting from a more sensitive detector than those studied before. We have lowered the detector baseline resolution from 3~eV to 0.6~eV by using tungsten TES films with lower $T_c$, which allows us to operate at lower bias voltages while retaining single-electron resolution. We were able to make a constrained measurement of overall phonon collection efficiency at 45\% (see \ref{app:PCE}), showing that this detector design is very close to the theoretical limit for phonon-to-quasiparticle conversion of $\sim$60\%~\cite{Guruswamy_2014}. Finally, we were able to show excellent agreement between the idealized TES noise model and data using only one free parameter, $G_{\text{inferred}}$. For the CUTE data, $G_{\text{inferred}} = 80.7$~pW/K was within 70\% of the bias power estimated value of $\sim$46.6~pW/K from Eq.~\ref{eq:G} assuming zero parasitics. The excellent agreement between best-fit values for $G_{\text{inferred}}$ across facilities builds a strong case that experimental values for $G$ obtained from raw bias-power data alone are subject to otherwise unconstrained parasitic power sources.

\subsection{Charge Recombination in High E-Field}

Our demonstrated improvement in HVeV detector energy resolution compared to earlier work has enabled us to tighten constraints on photon-based calibration systematics, as summarized in Table~\ref{tab: ctii-fit}. The measured sub-eV detector resolution also enables us to begin resolving photon peaks on top of charge peaks in energy histograms. We are thus able to directly demonstrate the simultaneous charge/phonon measurement technique which allows HVeV detectors to regain event type discrimination otherwise lost when explicit charge sensors are excluded from the detector design. As shown in Fig.~\ref{fig:led_cal}, each peak in the histogram of the SLAC near-UV photon calibration data exhibits a sharp edge at low energy, reflective of the excellent detector energy resolution, and a tail on the higher-energy side, due to Poisson statistics of electron-hole pairs generated by a given photon. Refs~\cite{Wilson_2024, ponce2020-1, ponce2020-2} describe in detail how we separate these and other physical effects observed in HVeV data. 

Most notably, we have significantly advanced the understanding of conservation processes for phonon energy vs. charge energy in cryogenic detectors capable of full charge collection. Past HVeV detector results~\cite{ren21,ComptonSteps,Albakry_2022} showed a systematic discrepancy between 0~V and HV ($\sim$ 10 - 100~V) energy scales. 
One potential cause for this difference is that electron-hole pairs that are generated by an event get separated fully and do not recombine in the bulk. When they terminate their trajectories on the surface, they do not fall back to the Fermi energy, but get trapped in higher energy states.
Circumstantial evidence from experiments with contact-free detectors, where substrate charge-up effects are noted regularly, is consistent with this hypothesis~\cite{hong20}. 
Another possibility is that the charges do fall down to the Fermi energy on the surface, but in a different manner than a recombination in the bulk, yielding a different athermal phonon efficiency at the QET.

Our demonstrated sub-eV energy resolution enabled robust measurements of the total energy recovered from a single photon event. By identifying clear consistencies between data obtained at two different facilities, using two different photon sources, we showed that for every event, some photon energy is ``lost". The CUTE and SLAC HVeV data show the loss to be $\sim$0.9~eV. The band-gap energy of Si at cryogenic temperatures is 1.17~eV~\cite{ramanathan2020ionization}, suggesting that some of the photon energy that initially creates electron-hole pairs is not recovered or recovered with a reduced efficiency when the charges reach the detector faces. This observation is consistent with the likely presence of band-bending effects at the edges of the substrate, due to surface states at the Si-vacuum interface, or is perhaps pointing towards a different phonon spectrum released during surface trapping, yielding a lower QET efficiency. In either hypothesis, $E_{\text{loss}}$ originates from the recombination of charges freed during a photon interaction and are therefore consistent with the lack of an observed dependence of $E_{\text{loss}}$ on the LED wavelength. Comparing the collected photon energy with the ionization bandgap, our data indicate possible band-bending on the order of 0.2-0.4~eV. Studies with the SLAC detector operated over a wider range of bias voltages and LED wavelengths will be the scope of future work, where we plan to better constrain energy-loss physics models relevant to these detectors.

\subsection{Implications for Correlated Phonon Noise}
\label{sec:discussion_corr_noise}
\setlength{\extrarowheight}{2pt}
\begin{table*}[!b]
\caption{Volume and phonon collection efficiencies of various devices from different experimental setups}
\label{tab: correlated noise}
  \centering
  \begin{tabular}{ccc}
  \hline 
\textbf{Detector} & \textbf{Volume} & \textbf{Phonon efficiency}\\
 & \textbf{(mm$^3$)} & \textbf{($\epsilon$)}\\
 \hline 
This work @ SLAC & 10 $\times$ 10 $\times$ 4 & $\sim$ 61\%\\ 
This work @ CUTE & 10 $\times$ 10 $\times$ 4 & $\sim$ 45\%\\
TESSERACT (1 mm) & 10 $\times$ 10 $\times$ 1 & $\sim$ 30\%\\
TESSERACT (4 mm) & 10 $\times$ 10 $\times$ 4 & $\sim$ 30\%\\
\end{tabular}
\end{table*}

\begin{figure}[!t]
    \centering
    \includegraphics[scale = 0.5]{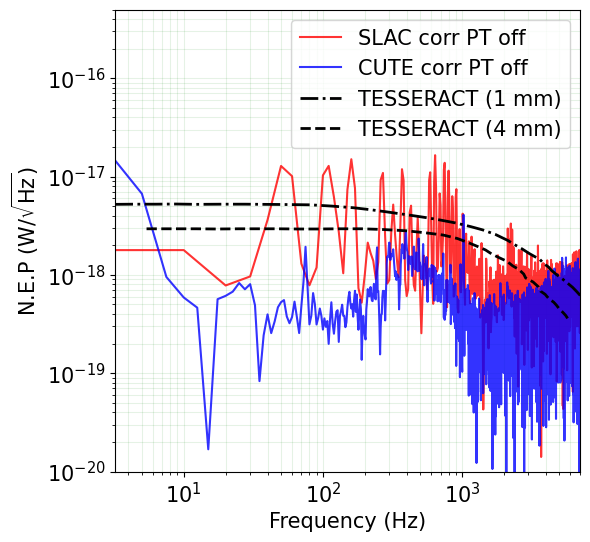}
    \caption{The correlated noise power spectra for CUTE and SLAC data for a single channel. The scaled correlated noise baseline for the 1 mm device from Ref~\cite{UCB25}  and 4 mm device from Ref~\cite{ucb2025_4mm} are shown for comparison. Here we only show the TESSERACT curve scaled to the CUTE efficiency. Scaling by the SLAC efficiency would shift both black curves up by $\sim$33\%. The peaks observed in the SLAC correlated noise can be attributed to 60 Hz pickup and associated harmonics. Note that the CUTE and SLAC data are partially transparent to show where they overlap in the high frequency region.}
    \label{fig:power_noise_discussion}
\end{figure}

As discussed in the introduction, the {\small TESSERACT} collaboration recently reported a resolution-limiting correlated power noise at low-frequencies, with magnitudes of $\sim$$9\cdot~10^{-19}$ W/$\sqrt{\rm Hz}$ ~\cite{UCB25}. Their reported baseline resolution of 375~meV using TES technology similar to ours and with comparable T$_c$, motivated a direct comparison between the fundamental noise performance of our respective detectors. 



Table~\ref{tab: correlated noise} compares the correlated NEP values observed by {\small TESSERACT} with those seen with our devices. As reported in Refs~\cite{UCB25, ucb2025_4mm}, the {\small TESSERACT} device exhibited excess correlated noise. By contrast, our detector noise performance (see Fig.~\ref{fig:power_noise}) is well-matched to the predicted TES model, aside from known pulse tube related low-frequency noise and environmental electronic noise peaks. We expect that some of the differences shown in Table~\ref{tab: correlated noise} can be attributed to detector design. For example, the thermal fluctuation noise described in Eq.~\ref{eq:TFN} scales with TES volume through $G\approx~5 {v}_{\text{TES}} \Sigma T_c^4$, where ${v}_{\text{TES}}$ is the tungsten TES volume, and $\Sigma$ is the TES electron-phonon coupling parameter. Assuming the same tungsten $T_c$ as that reported in Ref~\cite{UCB25}, we can estimate the {\small TESSERACT} thermal fluctuation noise by scaling our measured noise by the square root of the detector volume ratio between devices. In our detector, we have a total volume ${v}_{\text{TES}}\sim 6000~\mathrm{\upmu m}^3$. We estimate the {\small TESSERACT} TES volume based on the number of TESs and dimensions reported in Ref~\cite{caleb2022}: ${v}_{\text{TES}}\sim25\cdot(25\times2\times0.04)\sim 50~\mathrm{\upmu m}^3$. This suggests the NEP should be roughly $2.7\cdot 10^{-19}$~{W/$\sqrt{\rm Hz}$} for the {\small TESSERACT} detector. This result is about a factor of two higher than the modeled TES noise in their result~\cite{UCB25}, but about three times lower than their measured uncorrelated noise, which could point to sources other than intrinsic TES noise contributing to the reported {\small TESSERACT} results.


The comparison between CUTE and {\small TESSERACT}~\cite{UCB25} can be made more directly because they use similar \textit{CryoConcept} dilution refrigerators with vibration isolation, whereas SLAC uses an Oxford system without built-in, advanced vibration isolation. The measurement of the correlated noise is done with PT off data (as shown in Fig.~\ref{fig:power_noise_discussion}) to avoid contributions from PT vibrations. This is performed using the method described in \ref{app:appendixD}. We compare our resulting measurements of correlated noise to those of {\small TESSERACT}~\cite{UCB25, ucb2025_4mm}, scaled by detector volume and net QET phonon collection efficiency. 
The result of this exercise is shown in Fig.~\ref{fig:power_noise_discussion}, where an adaptation of the two correlated noise results reported by Ref~\cite{UCB25} and Ref~\cite{ucb2025_4mm} is shown in black. Here, we scaled the 1~mm {\small TESSERACT} device curve to the efficiency and volume of our device (NEP $\times$ 4~mm $\times \frac{\epsilon_{\text{CUTE}}}{\epsilon_{\text{TESSERACT}}}$). We also scaled the efficiency of the 4~mm device to that of our device (NEP $\times \frac{\epsilon_{\text{CUTE}}}{\epsilon_{\text{TESSERACT}}}$). The volume and efficiency values used for this scaling procedure are shown in Table~\ref{tab: correlated noise}.


Rather than a conclusive signature of excess correlated noise, we see two different trends in the data. The CUTE noise data does not show an excess above the white-noise baseline, aside from a peak at 1 kHz. The SLAC data shows a fairly flat baseline with some correlated noise peaks that exceed the {\small TESSERACT} baseline. Thus, if we assume a volume-scaled substrate-driven source underlying the excess correlated noise reported by {\small TESSERACT}, our measurements tentatively disfavor the hypothesis that silicon detectors exhibit a common, intrinsic, time-independent correlated phonon noise. 
Future investigations are required to clarify the sources of the correlated noise. We note better vibration mitigation will be a key challenge to such studies.

In summary, the demonstrated sub-eV energy resolution of our newest HVeV detectors represents a substantial advancement over prior implementations. We explored mechanisms of energy loss in Si detectors and measured a consistent loss of 0.9~eV per charge excitation between two facilities. Noise analyses indicate performance consistent with fundamental TES limits and no strong evidence for the correlated phonon noise reported in Ref~\cite{UCB25, ucb2025_4mm}. Together, these results establish a foundation for precise modeling of energy loss and noise processes in next-generation cryogenic detectors.

\section*{Acknowledgements}
\label{sec:acknowledgement}

We thank Prof. Belina von Krosigk for her support and for providing funding for part of the instrumentation used in this work.
The authors would like to thank the SuperCDMS collaboration for providing the DCRCs, SQUIDs and data acquisition software used in this work. These detectors were fabricated at Texas A\&M university with support provided by the Mitchel Institute. Funding and support were received from the Marie-Curie program, under contract No. 101104484.
This research was enabled in part by support provided by Compute Ontario (computeontario.ca) and the Digital Research Alliance of Canada (alliancecan.ca), NSERC Canada, the Canada First Excellence Research Fund, the Arthur B. McDonald Institute (Canada) and the Connaught Fund at University of Toronto. Finally, we thank SNOLAB and its staff for their support and use of the CUTE facility.
\printcredits

\bibliographystyle{elsarticle-num}

\bibliography{refs}

@article{albakry25,
    title = {{Light dark matter constraints from SuperCDMS HVeV detectors operated underground with an anticoincidence event selection}},
    author = {Albakry, M. F. and Alkhatib, I. and Alonso-Gonz\'alez, D. and Amaral, D. W. P. and Anczarski, J. and Aralis, T. and Aramaki, T. and Arnquist, I. J. and Ataee Langroudy, I. and Azadbakht, E. and others},
    journal = {Phys. Rev. D},
    volume = {111},
    issue = {1},
    pages = {012006},
    numpages = {9},
    year = {2025},
    month = {Jan},
    publisher = {American Physical Society},
    doi = {10.1103/PhysRevD.111.012006},
}

@ARTICLE{camus18,
    author = {{Camus}, Ph. and {Cazes}, A. and {Dastgheibi-Fard}, A. and {Dering}, K. and {Gerbier}, G. and {Rau}, W. and {Scorza}, S. and {Zhang}, X.},
    title = {{CUTE: A Low Background Facility for Testing Cryogenic Dark Matter Detectors}},
    journal = {J. Low Temp. Phys.},
    year = 2018,
    month = dec,
    volume = {193},
    number = {5-6},
    pages = {813-818},
    doi = {10.1007/s10909-018-2014-0},
    adsurl = {https://ui.adsabs.harvard.edu/abs/2018JLTP..193..813C}
}

@INPROCEEDINGS{hansen10,
    author={Hansen, Sten and DeJongh, Fritz and Hall, Jeter and Hines, Bruce A. and Huber, Martin E. and Kiper, Terry and Mandic, Vuk and Rau, Wolfgang and Saab, Tarek and Seitz, Dennis and Sundqvist, Kyle},
    booktitle={IEEE Nuclear Science Symposium \& Medical Imaging Conference}, 
    title={The Cryogenic Dark Matter Search test stand warm electronics card}, 
    year={2010},
    volume={},
    number={},
    pages={1392-1395},
    doi={10.1109/NSSMIC.2010.5874000}
}

@article{hong20,
    title={Single electron–hole pair sensitive silicon detector with surface event discrimination},
    volume={963},
    ISSN={0168-9002},
    DOI={10.1016/j.nima.2020.163757},
    journal={Nucl. Instrum. Methods A},
    publisher={Elsevier BV},
    author={Hong, Ziqing and Ren, Runze and Kurinsky, Noah and Figueroa-Feliciano, Enectali and Wills, Lise and Ganjam, Suhas and Mahapatra, Rupak and Mirabolfathi, Nader and Nebolsky, Brian and Pinckney, H. Douglas and others},
    year={2020},
    month=may, pages={163757}
}

@article{huber01,
    title={{DC SQUID series array amplifiers with 120 MHz bandwidth (corrected)}},
    author={Huber, Martin E and Neil, Patricia A and Benson, Robert G and Burns, Deborah A and Corey, AF and Flynn, Christopher S and Kitaygorodskaya, Yevgeniya and Massihzadeh, Omid and Martinis, John M and Hilton, GC},
    journal={IEEE Trans. Appl. Supercond.},
    volume={11},
    number={2},
    pages={4048--4053},
    year={2001},
    publisher={IEEE},
    doi={10.1109/77.919577},
}

@ARTICLE{irwin95,
    author = {{Irwin}, K.~D. and {Nam}, S.~W. and {Cabrera}, B. and {Chugg}, B. and {Young}, B.~A.},
    title = "{A quasiparticle-trap-assisted transition-edge sensor for phonon-mediated particle detection}",
    journal = {Rev. Sci. Instrum.},
    year = 1995,
    month = nov,
    volume = {66},
    number = {11},
    pages = {5322-5326},
    doi = {10.1063/1.1146105},
    adsurl = {https://ui.adsabs.harvard.edu/abs/1995RScI...66.5322I}
}

@ARTICLE{ren21,
    author = {{Ren}, R. and {Bathurst}, C. and {Chang}, Y.~Y. and {Chen}, R. and {Fink}, C.~W. and {Hong}, Z. and {Kurinsky}, N.~A. and {Mast}, N. and {Mishra}, N. and {Novati}, V. and others},
    title = {{Design and characterization of a phonon-mediated cryogenic particle detector with an eV-scale threshold and 100 keV-scale dynamic range}},
    journal = {Phys. Rev. D},
    year = 2021,
    month = aug,
    volume = {104},
    number = {3},
    eid = {032010},
    pages = {032010},
    doi = {10.1103/PhysRevD.104.032010},
}

@article{Camus_2024,
   title={{CUTE: A Cryogenic Underground TEst facility at SNOLAB}},
   volume={11},
   ISSN={2296-424X},
   DOI={10.3389/fphy.2023.1319879},
   journal={Front. Phys.},
   publisher={Frontiers Media SA},
   author={Camus, Philippe and Corbett, Jonathan and Crawford, Sean and Dering, Koby and Fascione, Eleanor and Gerbier, Gilles and Germond, Richard and Ghaith, Muad and Hall, Jeter and Hong, Ziqing and others},
   year={2024},
   month=jan }

@article{Romani_2018,
   title={Thermal detection of single e-h pairs in a biased silicon crystal detector},
    volume = {112},
    number = {4},
    pages = {043501},
    year = {2018},
    month = {01},
   ISSN={1077-3118},
   DOI={10.1063/1.5010699},
   journal={Appl. Phys. Lett.},
   publisher={AIP Publishing},
   author={Romani, R. K. and Brink, P. L. and Cabrera, B. and Cherry, M. and Howarth, T. and Kurinsky, N. and Moffatt, R. A. and Partridge, R. and Ponce, F. and Pyle, M. and others},
}

@article{Agnese_2018,
   title={{First Dark Matter Constraints from a SuperCDMS Single-Charge Sensitive Detector}},
   volume={121},
   ISSN={1079-7114},
   DOI={10.1103/physrevlett.121.051301},
   number={5},
   journal={Phys. Rev. Lett.},
   publisher={American Physical Society (APS)},
   author={Agnese, R. and Aralis, T. and Aramaki, T. and Arnquist, I. J. and Azadbakht, E. and Baker, W. and Banik, S. and Barker, D. and Bauer, D. A. and Binder, T. and others},
   year={2018},
   month=aug }

@article{Amaral_2020,
   title={{Constraints on low-mass, relic dark matter candidates from a surface-operated SuperCDMS single-charge sensitive detector}},
   volume={102},
   ISSN={2470-0029},
   DOI={10.1103/physrevd.102.091101},
   number={9},
   journal={Phys. Rev. D},
   publisher={American Physical Society (APS)},
   author={Amaral, D. W. and Aralis, T. and Aramaki, T. and Arnquist, I. J. and Azadbakht, E. and Banik, S. and Barker, D. and Bathurst, C. and Bauer, D. A. and Bezerra, L. V. S. and others},
   year={2020},
   month=nov }

@misc{HVeVR4,
      title={{Search for low-mass electron-recoil dark matter using a single-charge sensitive SuperCDMS-HVeV Detector}}, 
      collaboration={SuperCDMS Collaboration},
      author={M. F. Albakry and I. Alkhatib and D. Alonso-González and J. Anczarski and T. Aralis and T. Aramaki and I. Ataee Langroudy and C. Bathurst and R. Bhattacharyya and A. J. Biffl and others},
      year={2025},
      eprint={2509.03608},
      archivePrefix={arXiv},
      primaryClass={hep-ex},
}

@article{Ponce_2020,
   title={{Measuring the impact ionization and charge trapping probabilities in SuperCDMS HVeV phonon sensing detectors}},
   volume={101},
   ISSN={2470-0029},
   DOI={10.1103/physrevd.101.031101},
   number={3},
   journal={Phys. Rev. D},
   publisher={American Physical Society (APS)},
   author={Ponce, F. and Stanford, C. and Yellin, S. and Page, W. and Fink, C. and Pyle, M. and Sadoulet, B. and Serfass, B. and Watkins, S. L. and Brink, P. L. and others},
   year={2020},
   month=feb }

@article{Albakry_2022,
   title={{Investigating the sources of low-energy events in a SuperCDMS-HVeV detector}},
   volume={105},
   ISSN={2470-0029},
   DOI={10.1103/physrevd.105.112006},
   number={11},
   journal={Phys. Rev. D},
   publisher={American Physical Society (APS)},
   author={Albakry, M. F. and Alkhatib, I. and Amaral, D. W. P. and Aralis, T. and Aramaki, T. and Arnquist, I. J. and Ataee Langroudy, I. and Azadbakht, E. and Banik, S. and Bathurst, C. and others},
   year={2022},
   month=jun }

@article{Albakry_2023,
   title={{First Measurement of the Nuclear-Recoil Ionization Yield in Silicon at 100 eV}},
   volume={131},
   ISSN={1079-7114},
   DOI={10.1103/physrevlett.131.091801},
   number={9},
   journal={Phys. Rev. Lett.},
   publisher={American Physical Society (APS)},
   author={Albakry, M. F. and Alkhatib, I. and Alonso, D. and Amaral, D. W. P. and Aralis, T. and Aramaki, T. and Arnquist, I. J. and Ataee Langroudy, I. and Azadbakht, E. and Banik, S. and others},
   year={2023},
   month=aug }

@article{ComptonSteps,
  title = {{Low-energy calibration of SuperCDMS HVeV cryogenic silicon calorimeters using Compton steps}},
  author = {Albakry, M. F. and Alkhatib, I. and Alonso-Gonz\'alez, D. and Amaral, D. W. P. and Anczarski, J. and Aralis, T. and Aramaki, T. and Ataee Langroudy, I. and Bathurst, C. and Bhattacharyya, R. and others},
  collaboration = {SuperCDMS},
  journal = {Phys. Rev. D},
  volume = {112},
  issue = {9},
  pages = {092014},
  numpages = {14},
  year = {2025},
  month = {Nov},
  publisher = {American Physical Society},
  doi = {10.1103/jj7w-gkgg}
}

@article{Wilson_2024,
   title={Improved modeling of detector response effects in phonon-based crystal detectors used for dark matter searches},
   volume={109},
   ISSN={2470-0029},
   DOI={10.1103/physrevd.109.112018},
   number={11},
   journal={Phys. Rev. D},
   publisher={American Physical Society (APS)},
   author={Wilson, M. J. and Zaytsev, A. and von Krosigk, B. and Alkhatib, I. and Buchanan, M. and Chen, R. and Diamond, M. D. and Figueroa-Feliciano, E. and Harms, S. A. S. and Hong, Z. and others},
   year={2024},
   month=jun }

@article{Kelsey_2023,
   title={{G4CMP: Condensed matter physics simulation using the Geant4 toolkit}},
   volume={1055},
   ISSN={0168-9002},
   DOI={10.1016/j.nima.2023.168473},
   journal={Nucl. Instrum. Methods A},
   publisher={Elsevier BV},
   author={Kelsey, M.H. and Agnese, R. and Alam, Y.F. and Langroudy, I. Ataee and Azadbakht, E. and Brandt, D. and Bunker, R. and Cabrera, B. and Chang, Y.-Y. and others},
   year={2023},
   month=oct, pages={168473} }

@article{ZurekReview,
   author = "Zurek, Kathryn M.",
   title = "{Dark Matter Candidates of a Very Low Mass}", 
   journal= "Annu. Rev. Nucl. Part. Sci.",
   year = "2024",
   volume = "74",
   pages = "287-319",
   doi = "https://doi.org/10.1146/annurev-nucl-101918-023542",
   publisher = "Annual Reviews",
   issn = "1545-4134",
   type = "Journal Article"
  }

@article{Kahn_2022,
   title={Searches for light dark matter using condensed matter systems},
   volume={85},
   ISSN={1361-6633},
   DOI={10.1088/1361-6633/ac5f63},
   number={6},
   journal={Rep. Prog. Phys.},
   publisher={IOP Publishing},
   author={Kahn, Yonatan and Lin, Tongyan},
   year={2022},
   month=may, pages={066901} }

@book{Murayama_2023,
   title={Exploring the Quantum Universe: Pathways to Innovation and Discovery in Particle Physics},
   DOI={10.2172/2368847},
   institution={Office of Scientific and Technical Information (OSTI)},
   author={Murayama, Hitoshi and Asai, Shoji and Heeger, Karsten and Ballarino, Amalia and Bose, Tulika and Cranmer, Kyle and Cyr-Racine, Francis-Yan and Demers, Sarah and Geddes, Cameron and Gershtein, Yuri and others},
   year={2023},
   month=jun }

@article{Augier_2023,
 title = {{Characterization of mini-CryoCube detectors from the RICOCHET experiment commissioning at the Institut Laue-Langevin}},
  author = {Armatol, A. and Augier, C. and Bailly-Salins, L. and Baulieu, G. and Berg\'e, L. and Billard, J. and Bl\'e, J. and Bres, G. and Bret, J-.L. and Broniatowski, A. and others},
  collaboration = {RICOCHET},
  journal = {Phys. Rev. D},
  volume = {112},
  issue = {11},
  pages = {112019},
  numpages = {16},
  year = {2025},
  month = {Dec},
  publisher = {American Physical Society},
  doi = {10.1103/7xy6-jq3c}
}

@article{nucleus22,
  title = {{Commissioning of the NUCLEUS Experiment at the Technical University of Munich}},
  author = {Abele, H. and Angloher, G. and Arnold, B. and Atzori Corona, M. and Bento, A. and Bossio, E. and Buchsteiner, F. and Burkhart, J. and Cappella, F. and Cappelli, M. and others},
  collaboration = {NUCLEUS},
  journal = {Phys. Rev. D},
  volume = {112},
  issue = {7},
  pages = {072013},
  numpages = {21},
  year = {2025},
  month = {Oct},
  publisher = {American Physical Society},
  doi = {10.1103/c95p-8kh2}
}

@article{UCB25,
   title={Low energy backgrounds and excess noise in a two-channel low-threshold calorimeter},
   volume={126},
   ISSN={1077-3118},
   DOI={10.1063/5.0247343},
   number={10},
   journal={Appl. Phys. Lett.},
   publisher={AIP Publishing},
   author={Anthony-Petersen, R. and Chang, C. L. and Chang, Y.-Y. and Chaplinsky, L. and Fink, C. W. and Garcia-Sciveres, M. and Guo, W. and Hertel, S. A. and Li, X. and Lin, J. and others},
   year={2025},
   month=mar }

@article{Adari_2022,
    title={{EXCESS workshop: Descriptions of rising low-energy spectra}},
    ISSN={2666-4003},
    DOI={10.21468/scipostphysproc.9.001},
    volume={9},
    pages={001},
    journal={SciPost Phys. Proc.},
    publisher={Stichting SciPost},
    author={Adari, Prakruth and Aguilar-Arevalo, Alexis A. and Amidei, Dante and Angloher, G. and Armengaud, Eric and Augier, C. and Balogh, Levente and Banik, S. and Baxter, David and Beaufort, C. and others},
    year={2022},
    month=aug }

@article{EXCESS2025,
   author = "Baxter, Daniel and Essig, Rouven and Hochberg, Yonit and Kaznacheeva, Margarita and von Krosigk, Belina and Reindl, Florian and Romani, Roger K. and Wagner, Felix",
   title = "Low-Energy Backgrounds in Solid-State Phonon and Charge Detectors", 
   journal= "Annual Review of Nuclear and Particle Science",
   year = {2025},
   volume = {75},
   number = {},
   pages = "301--326",
   DOI = "10.1146/annurev-nucl-121423-100849",
   publisher = "Annual Reviews",
   ISSN = {1545-4134},
   type = "Journal Article",
  }

@Inbook{Irwin2005,
    author="Irwin, K.D.
    and Hilton, G.C.",
    editor="Enss, Christian",
    chapter="{Transition-Edge Sensors}",
    Title="Cryogenic Particle Detection",
    year="2005",
    publisher="Springer Berlin Heidelberg",
    address="Berlin, Heidelberg",
    pages="63--150",
    isbn="978-3-540-31478-3",
    doi="10.1007/10933596_3"
}

@article{Fink_2020,
   title={{Characterizing TES power noise for future single optical-phonon and infrared-photon detectors}},
   volume={10},
   ISSN={2158-3226},
   DOI={10.1063/5.0011130},
   number={8},
   journal={AIP Adv.},
   publisher={AIP Publishing},
   author={Fink, C. W. and Watkins, S. L. and Aramaki, T. and Brink, P. L. and Ganjam, S. and Hines, B. A. and Huber, M. E. and Kurinsky, N. A. and Mahapatra, R. and Mirabolfathi, N. and others},
   year={2020},
   month=aug }

@article{Guruswamy_2014,
   title={Quasiparticle generation efficiency in superconducting thin films},
   volume={27},
   ISSN={1361-6668},
   DOI={10.1088/0953-2048/27/5/055012},
   number={5},
   journal={Supercond. Sci. Technol.},
   publisher={IOP Publishing},
   author={Guruswamy, T and Goldie, D J and Withington, S},
   year={2014},
   month=mar, pages={055012} }

@article{ramanathan2020ionization,
    title={{Ionization yield in silicon for eV-scale electron-recoil processes}},
    author={Ramanathan, Karthik and Kurinsky, Noah},
    journal={Phys. Rev. D},
    volume={102},
    number={6},
    pages={063026},
    year={2020},
    publisher={APS},
    doi = {10.1103/PhysRevD.102.063026}
}

@article{ucb2025_4mm,
    author = {Chang, C. L. and Chang, Y.-Y. and Garcia-Sciveres, M. and Guo, W. and Hertel, S. A. and Li, X. and Lin, J. and Lisovenko, M. and Mahapatra, R. and Matava, W. and others},
    title = {Spontaneous generation of athermal phonon bursts within bulk silicon causing excess noise, low energy background events, and quasiparticle poisoning in superconducting sensors},
    journal = {Appl. Phys. Lett.},
    volume = {127},
    collaboration ={TESSERACT},
    number = {26},
    pages = {263502},
    year = {2025},
    issn = {0003-6951},
    doi = {10.1063/5.0281876}
}

@article{qrocodile,
  title = {{First Sub-MeV Dark Matter Search with the QROCODILE Experiment Using Superconducting Nanowire Single-Photon Detectors}},
  author = {Baudis, Laura and Bismark, Alexander and Brugger, Noah and Capelli, Chiara and Charaev, Ilya and Garc\'{\i}a, Jose Cuenca and Hadas, Guy Daniel and Hochberg, Yonit and Hohmann, Judith K. and Kavner, Alexander and others},
  journal = {Phys. Rev. Lett.},
  volume = {135},
  issue = {8},
  pages = {081002},
  numpages = {8},
  year = {2025},
  month = {Aug},
  publisher = {American Physical Society},
  doi = {10.1103/4hb6-f6jl},
}

@article{TESSERACT,
    title = {{Transition Edge Sensors with Sub-eV Resolution And Cryogenic Targets (TESSERACT) at the underground laboratory of Modane (LSM)}},
    journal = {Nucl. Phys. B},
    volume = {1003},
    pages = {116465},
    year = {2024},
    note = {Special Issue of Nobel Symposium 182 on Dark Matter},
    issn = {0550-3213},
    doi = {https://doi.org/10.1016/j.nuclphysb.2024.116465},
    author = {J. Billard and J. Gascon and S. Marnieros and S. Scorza}
}

@misc{caleb_w_fink_2022_5903353,
  author       = {Caleb W Fink and
                  Samuel L Watkins},
  title        = {{QETpy}},
  month        = jan,
  year         = 2022,
  publisher    = {Zenodo},
  version      = {1.3.2},
  doi          = {10.5281/zenodo.5903353},
}

@article{maasilta_2012,
    author = {Maasilta, I. J. },
    title = {{Complex impedance, responsivity and noise of transition-edge sensors: Analytical solutions for two- and three-block thermal models}},
    journal = {AIP Advances},
    year = {2012},
    volume = {2},
    pages = { 042110}
}

@phdthesis{caleb2022,
    author = {Fink, C.},
    title = {{A Gram-Scale low-Tc Low-Surface-Coverage Athermal-Phonon Sensitive Dark Matter Detector}},
    school = {UC Berkeley},
    year = {2022}
}

@article{ponce2020-1,
    author = {Ponce, F. and Page, W. and Brink, P.L. and Cabrera, B. and Cherry, M. and Fink, C. and Kurinsky, N. and Partridge, R. and Pyle, M. and Sadoulet, B. and others},
    title = {{Modeling of Impact Ionization and Charge Trapping in SuperCDMS HVeV Detectors}},
    journal = {J. Low Temp. Phys.},
    year = {2020},
    volume = {199},
    pages = {598-605},
    doi = {10.1007/s10909-020-02349-x}
}

@article{ponce2020-2,
    title = {{Measuring the impact ionization and charge trapping probabilities in SuperCDMS HVeV phonon sensing detectors}},
    author = {Ponce, F. and Stanford, C. and Yellin, S. and Page, W. and Fink, C. and Pyle, M. and Sadoulet, B. and Serfass, B. and Watkins, S. L. and Brink, P. L. and others},
    journal = {Phys. Rev. D},
    volume = {101},
    issue = {3},
    pages = {031101},
    numpages = {5},
    year = {2020},
    month = {Feb},
    publisher = {American Physical Society},
    doi = {10.1103/PhysRevD.101.031101}
}

@article{ALLISON2016186,
    title = {{Recent developments in Geant4}},
    journal = {Nucl. Instrum. Methods A},
    volume = {835},
    pages = {186-225},
    year = {2016},
    issn = {0168-9002},
    doi = {10.1016/j.nima.2016.06.125},
    author = {J. Allison and K. Amako and J. Apostolakis and P. Arce and M. Asai and T. Aso and E. Bagli and A. Bagulya and S. Banerjee and G. Barrand and others}
}

@article{allison2006,
    author = {Allison, J. and Amako, K. and Apostolakis, J. and Araujo, H. and Arce Dubois, P. and Asai, M. and Barrand, G. and Capra, R. and Chauvie, S. and Chytracek, R. and others},
    title = {{Geant4 Developments and Applications}},
    journal = {IEEE Trans. Nucl. Sci. },
    year = {2006},
    volume = {53},
    pages = {270 - 278}
}

@article{agostinelli2003,
    author = {S. Agostinelli and J. Allison and K. Amako and J. Apostolakis and H. Araujo and P. Arce and M. Asai and D. Axen and S. Banerjee and G. Barrand and others},
    title = {{Geant4 - A Simulation Toolkit}},
    journal = {Nucl. Instrum. Meth. A},
    year = {2003},
    volume = {506},
    pages = {250 - 303}
}

@incollection{TESBible,
  title={Transition-edge sensors},
  author={Irwin, Kent D and Hilton, Gene C},
  booktitle={Cryogenic particle detection},
  pages={63--150},
  year={2005},
  publisher={Springer}
}

@article{bartrud,
    author = {Bratrud, G. and Chang, C.L. and Chen, R. and Cudmore, E. and Figueroa-Feliciano, E. and Hong, Z. and Kennard, K. T. and Lewis, S. and Lisovenko, M. and Mateo, L. O. and others},
    title = {{First demonstration of a TES based cryogenic Li$_2$MoO$_4$ detector for neutrinoless double beta decay search}},
    journal = {Eur. Phys. J. C},
    volume = {85}, 
    pages = {118}, 
    year = {2025},
    doi = {10.1140/epjc/s10052-025-13844-4} 
}

@article{CRESST_Res,
  title = {First observation of single photons in a {CRESST} detector and new dark matter exclusion limits},
  author = {Angloher, G. and Banik, S. and Benato, G. and Bento, A. and Bertolini, A. and Breier, R. and Bucci, C. and Burkhart, J. and Canonica, L. and D'Addabbo, A. and others},
  collaboration = {CRESST Collaboration},
  journal = {Phys. Rev. D},
  volume = {110},
  issue = {8},
  pages = {083038},
  numpages = {11},
  year = {2024},
  month = {Oct},
  publisher = {American Physical Society},
  doi = {10.1103/PhysRevD.110.083038},
}


\appendix

\renewcommand{\thesection}{Appendix \Alph{section}}

\section{Phonon Collection Efficiency}\label{app:PCE}
The phonon collection efficiency refers to the fraction of phonon energy that is absorbed by the TES, and can be thought of as the QET total quantum efficiency, as it includes both phonon and internal energy losses. The total energy efficiency is the product of the following four process:
\begin{enumerate}
    \item Phonon collection efficiency 
    \item Phonon to quasiparticle conversion efficiency
    \item QP collection efficiency
    \item Trapped QP to TES thermal energy conversion efficiency
\end{enumerate}
The absorbed energy is computed as the integral of the power pulse over the time it takes to return to baseline $T$ as
\begin{align}
      E_\mathrm{abs} &= \int_{0}^{T} \Delta P dt\\
      &= \int_{0}^{T} \left[ \left(V_b - 2I_0 R_l\right)\Delta I(t) - \Delta I(t)^2 R_l \right] dt,
    \label{eq:4.3.1}
\end{align}
where $\Delta I$ refers to the change in TES current from a small change in temperature, $R_l = R_p + R_{\text{shunt}}$, and $V_b$ is the bias voltage. The measured energy from both channels are summed to obtain the total measured energy for each event, $E_{\text{abs}-\text{tot}}$. The net phonon collection efficiency is estimated as
\begin{equation}
    \epsilon = \frac{E_{\text{abs}-\text{tot}}}{E_\mathrm{expected}} \times 100\%,
\end{equation}
where $E_\mathrm{expected}$ is the expected energy of the event given by Eq.~\ref{eq:phononEnergy}.

\begin{figure*}[!h]
\centering
    \begin{subfigure}[b]{0.5\textwidth}
        \includegraphics[width=.8\textwidth]{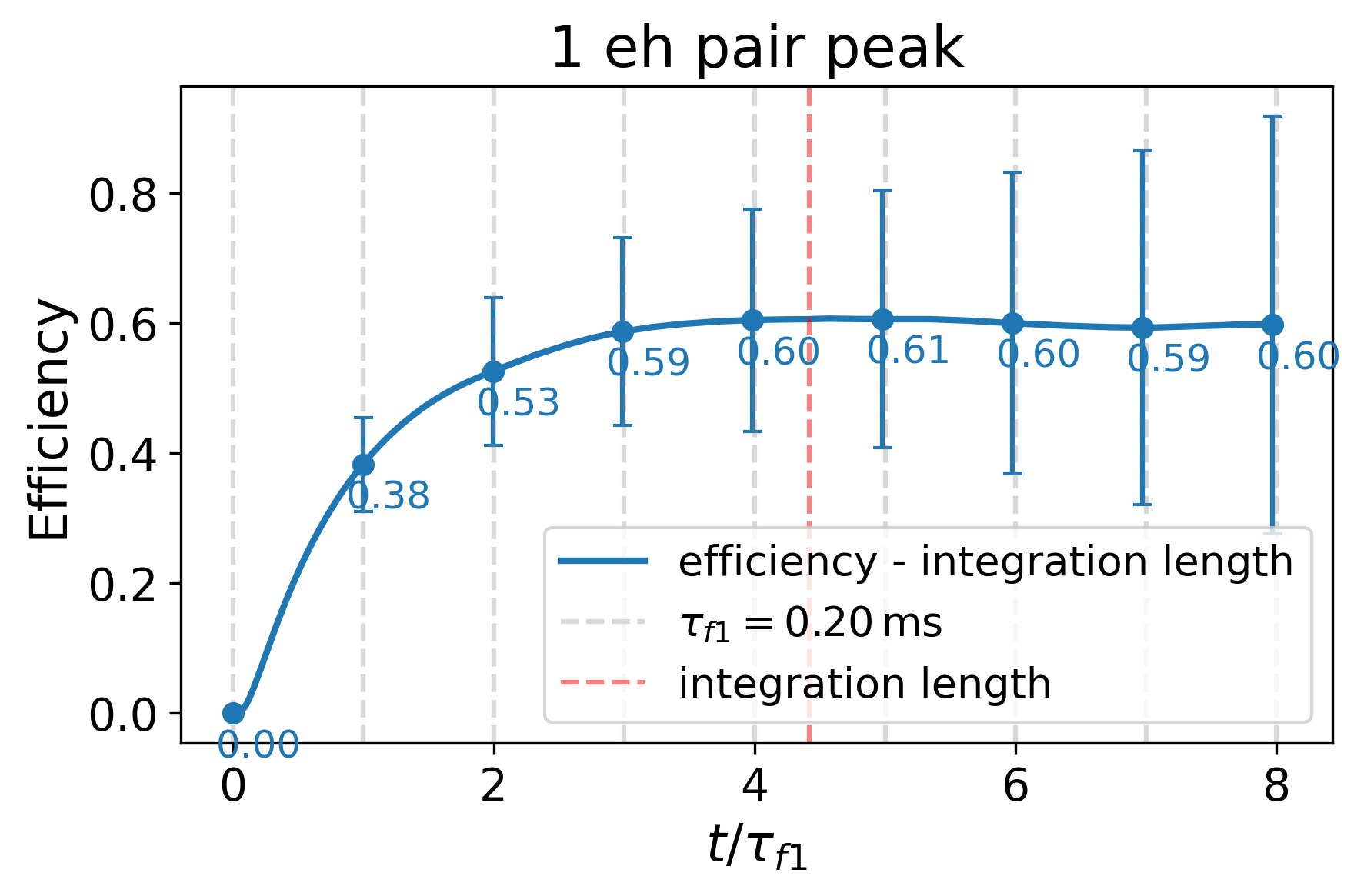}
        \caption{SLAC}
         \label{fig:subfig1}
    \end{subfigure}
    \hspace{1cm}
    \begin{subfigure}[b]{0.4\textwidth}
        \includegraphics[width=1.\textwidth]{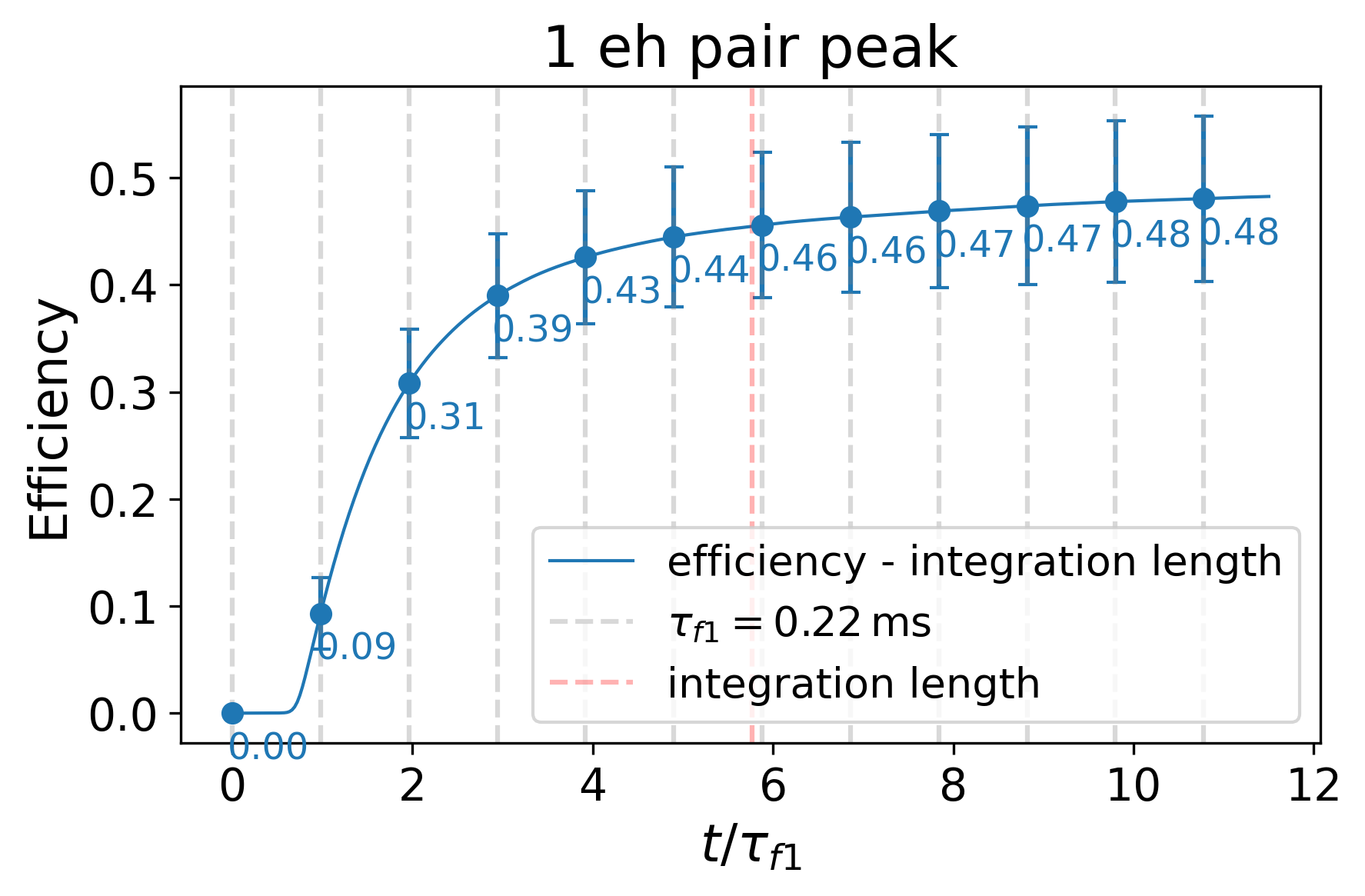}
        \caption{CUTE}
        \label{fig:subfig2}
    \end{subfigure}
    \caption{The phonon collection efficiency with 1$\sigma$ uncertainty  ($ \sqrt{\sigma_{stat}^2+\sigma_{sys}^2} $) as a function of integration range. The x axis is time/fall time from the double exponential fit of the pulse template. \textbf{Left} shows the efficiency of the quantized 1e$^-$/h$^+$ peak from SLAC data. \textbf{Right} shows the efficiency of only the 1e$^-$/h$^+$ pair at CUTE.}
\label{fig:efficiency}
\end{figure*}

At SLAC, we calibrate the detector by illuminating it with mono-energetic photons from an LED to produce quantized single e$^-$/h$^+$ peaks in the energy spectrum (see Sec.~\ref{sec:phonon_calibration}). For data taken at CUTE, we use single e$^-$/h$^+$ events from background data in order to avoid contributions from cross-talk that is present in LED data, as discussed in \ref{app:crosstalk}. These peaks serve as known energy points for the efficiency calculation. The energy spectrum is produced by converting the current pulse in amperes to $E_\mathrm{abs}$ in eV using Eq.~\ref{eq:4.3.1}. The integration range is varied by fixing the starting point of each pulse and scanning over the tail of the pulse to determine the optimal cut off point. Figure~\ref{fig:efficiency} shows the phonon collection efficiency as a function of integration length shown in fractions of the pulse fall time. The efficiency begins to plateau beyond five fall times and therefore, we quote this value as the measured phonon collection efficiency for these devices.

Table~\ref{tab:efficiency} compares the collection efficiency and the statistical and systematic uncertainties for the first quantized peak between SLAC and CUTE. The integral is a less robust estimator in the presence of correlated noise as in the case of the SLAC data. Due to its limited constraining power, we use the SLAC data solely to establish an upper bound on the efficiency. In contrast, the CUTE data is not limited by noise, allowing us to constrain the phonon collection efficiency to approximately 45\%.

\begin{table}[htbp]
    \centering
    \caption{Phonon collection efficiency with statistical and systematic uncertainties for SLAC and CUTE from the first quantized peak.}
    \begin{tabular}{lccc}
        \toprule
        & Efficiency & Statistical & Systematic \\
        \midrule
        SLAC & 61\% & $\pm$ 16\% & $\pm$ 7\% \\
        CUTE & 45\% & $\pm$ 3\% & $\pm$ 6\% \\
        \bottomrule
    \end{tabular}
    
    \label{tab:efficiency}
\end{table}

\section{Triggering and Event Reconstruction}\label{app:Trigger}
Data were acquired as a continuous time-stream and triggered offline. For background data taken at CUTE, a threshold trigger was applied to time series after passing through a gaussian-derivative filter. When the threshold trigger activates, a peak search window with a width of 1024 samples ($\sim$6.5 ms) after the initial trigger is defined. The trigger point is then adjusted to the time at which the filtered time-series reaches a maximum within the window. To study noise conditions, we also triggered at a random time within each 500 ms window of the continuous time series. For calibration data, an external trigger was configured to trigger on the rising edge of the voltage signal supplied to the LED which enabled triggering on LED events with little to no observable signal. For each trigger, a 2048 sample ($\sim$13 ms) window centered on the trigger point is saved for additional analysis. 

\begin{figure}[b]
    \includegraphics[width=\columnwidth]{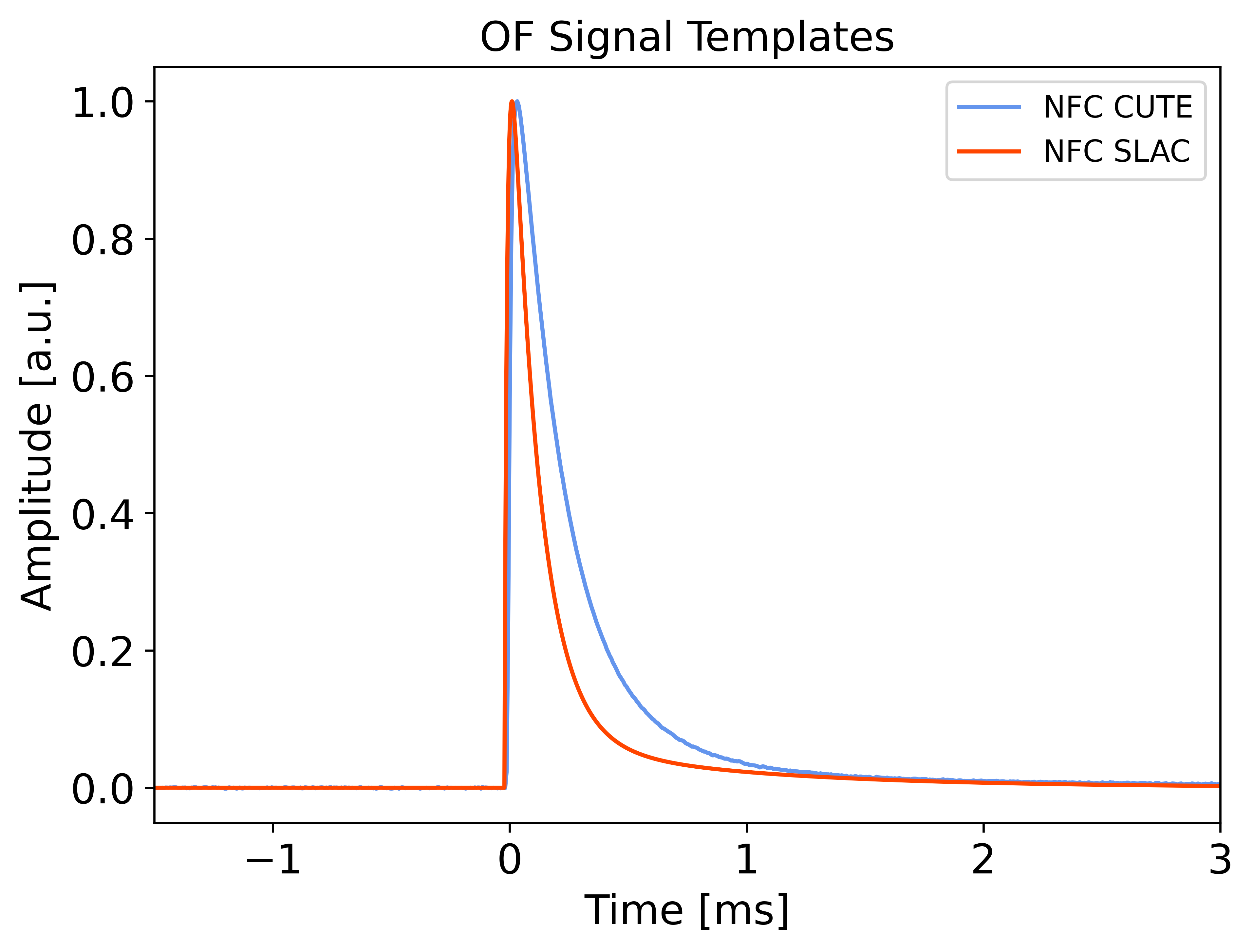}
    \caption{The input signal templates (for CUTE in blue and for SLAC in red) supplied at inputs for the OF. The templates are the same length as the analysis windows saved for each event trigger. Here we have zoomed in around the pulse to facilitate comparisons. Refer to \ref{sub_sec:tes_characterization} for a discussion on the difference in fall times between facilities.
    }\label{fig:templates}
\end{figure}

The amplitude of triggers is estimated using a frequency-domain optimum filter (OF) which takes as an input a noise PSD and a signal template. The noise PSD is calculated from noise triggers which have passed through strict selections based on their mean, slope, skew, and range (the difference in the maximum and minimum of the trace). At CUTE, data were processed with an initial template (referred to as a pre-template) which matched pulse shapes of similar detectors operated in the past. The signal templates were then constructed by averaging a sample of 1750 pulses which were selected from the 1st electron-hole-pair peak in background data according to the following criteria: 1) The events' $\chi^2$ comparing the pulse to the pre-template in frequency space are within 1$\sigma$ of the mean $\chi^2$, 2) the mean and slope of the baseline before the onset of the pulse must be within $3\sigma$ of the mean of the distribution of mean baselines.

For data taken at SLAC, data is saved as 52 ms long time-streams collected at 3-4 Hz. The sample frequency at SLAC was 4 times greater than at CUTE which enabled the calculation of higher-resolution noise PSDs for the OF processing and correlated noise studies without diminishing live time. For the calibration data, similar to the setup at CUTE, an external trigger was configured so that the trigger frequency matched the frequency of LED pulses. For each trigger, a 7000 sample ($\sim$11 ms) window with 1000 samples pre-trigger is saved for OF analysis. For noise studies, a randomly triggered window was saved for every time domain trace. 

The amplitudes of the pulses are determined with the same OF process used for CUTE data. At SLAC, the signal templates were generated by fitting a simple 2-pole pulse-shape (Eq. \ref{eq:pulse-eq}) to the low-pass filtered averaged traces as defined below
\begin{equation}
    I_\text{template}(t) = \begin{cases}
0 & \text{if } t \leq t_0\\
-e^{-(t-t_0)/\tau_1} + e^{-(t-t_0)/\tau_2}  & \text{if } t > t_0
\end{cases},
\label{eq:pulse-eq}
\end{equation}
where $t_0$ is the pre-trigger time and $\tau_1, \tau_2$ are the rise and fall time respectively. This is then normalized to have peak amplitude 1. The final templates used for event reconstruction are shown in Fig.~\ref{fig:templates}.

The OF minimizes the $\chi^2$ by fitting both the amplitude of the pulse and the time of the onset of the pulse which is limited to a small window around the initial trigger — the resulting amplitude is referred to as A$_{\rm OFL}$. However, the OF can also be constrained to a single parameter by fixing the trigger time so that the only fitting parameter is the amplitude of the pulse. In this case, the amplitude is referred to as A$_{\rm OF0}$. Fixing the arrival time allows for the zero-pulse amplitude distribution to remain gaussian, while a floating arrival time leads to a look-elsewhere effect that tends to push the zero-pulse amplitudes away from zero. Here, we primarily utilized $A_{\text{OF0}}$ in order to constrain baseline resolution and avoid threshold effects from the time-shifting OF.

\section{Cross-talk in CUTE Photon Calibration}\label{app:crosstalk}
In the experimental setup at CUTE, the calibration LEDs were mounted within the cryogenic detector payloads, which caused measurable cross-talk between the LED cabling and TES signal readout. The dominant cross-talk effects included: (1) A fast "spike" from the LED pulsing signal that immediately preceded the TES phonon signal, and (2) a much slower component, similar to a phonon pulse, that if left uncorrected would increase the estimated amplitude returned by the optimum filter.

To reduce the effect of the fast-component on the amplitude estimate, we calibrated using the amplitude with a fixed time delay relative to the trigger, A$_{\text{OF}0}$ (as opposed to A$_{\text{OFL}}$) since it was observed that the cross-talk ``spike'' could shift the best-fit time delay, which results in an underestimate in the pulse amplitude.  The amplitude of the latter component of the cross-talk was not observed to scale with the amplitude of the voltage pulse supplied to the LED. Assuming that the magnitude of the pulse component of cross-talk was independent of the energy deposited in the crystal substrate, we modeled its effect as a constant shift, A$_{\text{CT}}$, to the histogram of A$_{\text{OF}0}$ values for LED data. 

\begin{figure}[!t]
    \centering
    \includegraphics[width=\columnwidth]{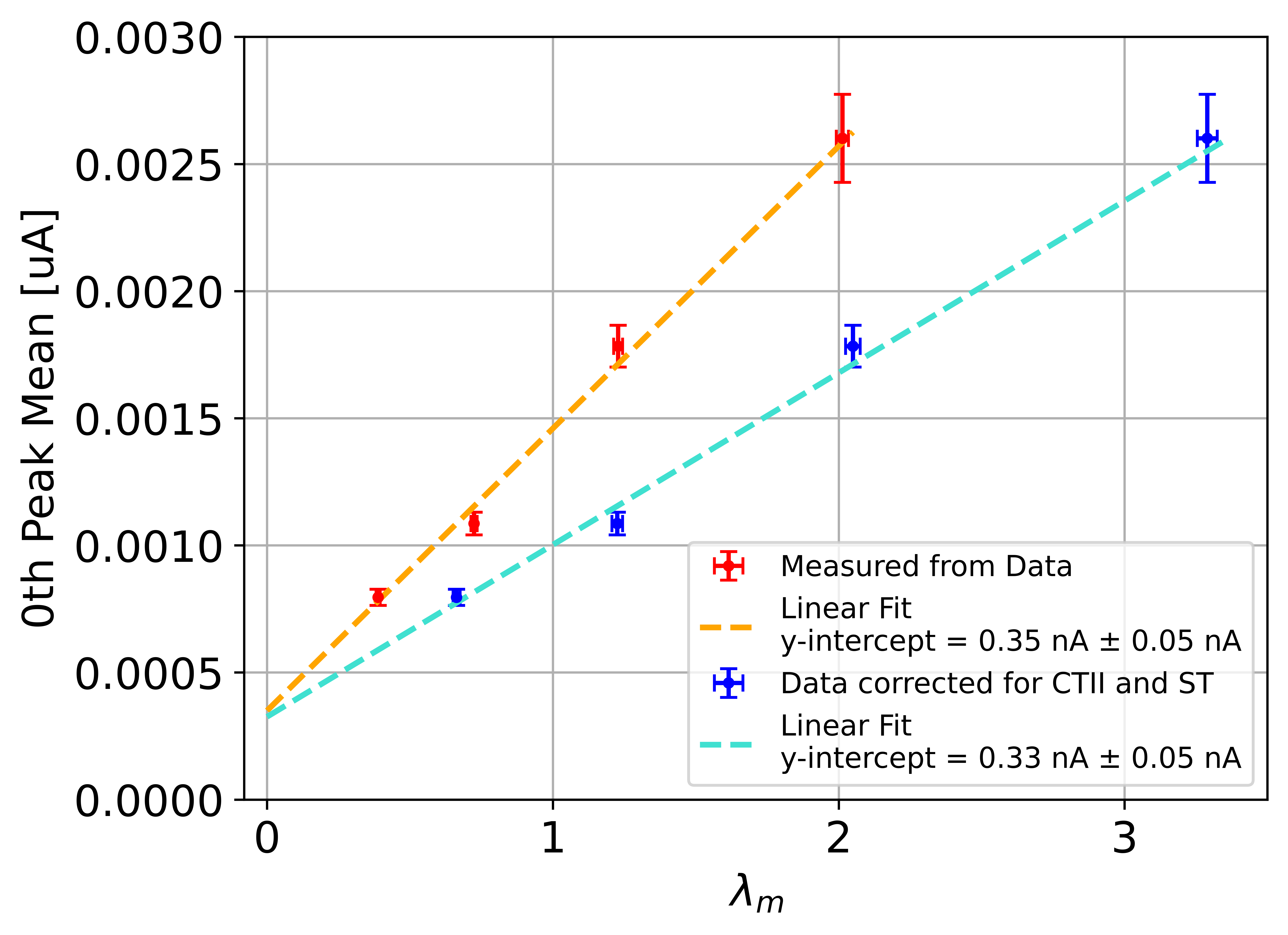}
    \caption{The data in red shows the mean of the 0th peak as a function of $\lambda_{m}$, the mean of the Poisson distribution describing the relative heights of the eh peaks resulting from a direct measurement to the data. The y-intercept of the corresponding linear fit (shown as the orange dashed line) is taken as an initial estimate for A$_{\text{CT}}$. The data plotted in blue accounts for the expected effect of the surface-trapping probability on $\lambda_{m}$. Here, the corresponding linear fit (shown as the turquoise dashed line) yields a slightly lower estimate for A$_{\text{CT}}$. 
    }\label{fig:Lam0th}
\end{figure}

The value of A$_{\text{CT}}$ was first estimated by observing the position of the 0th peak as a function of $\lambda_{m}$, the measured mean of the Poisson distribution that describes the observed heights of the e$^-$/h$^+$ peaks. Since surface-trapping and charge-trapping can shift events away from e$^-$/h$^+$ peaks, there is a subtle difference between $\lambda_m$ and the mean number of photons absorbed in the detector (referred to as $\lambda$) depending on the probability of both surface-trapping (P$_{\text{ST}}$) and charge-trapping (P$_{\text{CT}}$). Both cross-talk and surface trapped events shift the mean of the 0th peak to higher A$_{\text{OF}0}$. We separated these effects by characterizing the 0th peak mean as a function of $\lambda_{m}$ and extrapolating to $\lambda_{m} = 0$ at which point the only contribution to the 0th peak mean is A$_{\text{CT}}$. We observed a linear relationship (see the orange curve in Fig.~\ref{fig:Lam0th}) between the 0th peak mean and $\lambda_{m}$ thus A$_{\text{CT}}$ was taken to be the y-intercept of the best fit line.

To account for any systematic uncertainty introduced by cross-talk, we included A$_{\text{CT}}$ as a fitting parameter which would shift the detector response model along the x-axis. In accordance with our original estimate, we included a gaussian prior on A$_{\text{CT}}$ with a mean of 0.35 nA and standard deviation of 0.05 nA. The result of the fit gave A$_{\text{CT}}$ = 0.41 $\pm$ 0.02 nA which differed from our estimate by $\sim$1$\sigma$. However, as previously mentioned, the presence of charge trapping and surface trapping may affect the estimate of A$_{\text{CT}}$. We modeled the effect of a surface-trapping probability of 21.9\% (corresponding to the fit result) on $\lambda_{m}$ and repeated the linear fit which yielded an estimate for A$_{\text{CT}}$ of 0.32 $\pm$ 0.05 nA (see the data shown in blue in Fig. \ref{fig:Lam0th} and the corresponding linear fit). This corresponds to a $\sim$1.8$\sigma$ discrepancy between the result of the fit and our estimation of A$_{\text{CT}}$ thus we conclude the results are in agreement.

\section{Performance of the decorrelation algorithm}
\label{app:appendixD}

Consider a device with two channels (A, B) where the measured noise in each channel arises from a combination of its intrinsic noise ($\sigma_A(f),\;\sigma_B(f)$) and a common correlated noise source shared by both channels, $\sigma_C(f)$. In this case, naively calculating the PSD for a single channel would give:
\begin{equation}
    \hat{J}_i(f) = |A_i(f)|^2\sigma_C(f)^2 + \sigma_{i}(f)^2
    \label{eq:niave_j},
\end{equation}
where $A_i(f)$ is the coupling strength for the correlated noise in the respective channels. i.e. the total phonon collection efficiency. If instead we calculate the total covariance matrix in the frequency domain, $\Sigma(f)$ is given by the $2\times 2$ matrix:
\begin{equation*}
\begin{bmatrix} 
    \left|A_A(f)\right|^2\sigma_C(f)^2 + \sigma_{A}(f)^2 & A_A(f)A_B(f)^*\sigma_C(f)^2\\ A_B(f)A_A(f)^*\sigma_C(f)^2 & \left|A_B(f)\right|^2\sigma_C(f)^2 + \sigma_{B}(f)^2\\
\end{bmatrix}.
\end{equation*}
 Looking at the reciprocal of the diagonal elements of the inverse of the covariance matrix above, we get (for channel A)
 
\begin{align}
    \frac{1}{\left[\mathbf{\Sigma}(f) ^{-1}\right]_{A,A}}  = \sigma_{A}(f)^2 +\frac{\left|A_A(f)\right|^2 \sigma_C(f)^2 \sigma_{B}(f)^2}{\left|A_B(f)\right|^2\sigma_C(f)^2 +\sigma_{B}(f)^2}.
\end{align}

In the limit that channel B is dominated by correlated noise ($\left|A_B(f)\right|^2\sigma_C(f)^2 \gg \sigma_{B}(f)^2$), assuming $A_A = A_B$ we get 
\begin{align*}
     \frac{1}{\left[\mathbf{\Sigma}(f) ^{-1}\right]_{A,A}}=  \hat{J}_{uncorr}^{A,A}(f)= \sigma_{A}(f)^2 + \sigma_{B}(f)^2,\\
     \textrm{(Correlated noise dominated)}
\end{align*}
which is to say, that in the limit that we are dominated by correlated noise, this provides an estimator of the uncorrelated component of the noise. 

Alternatively, in the limit where the intrinsic noise is dominant, we get

\begin{align*}
     \frac{1}{\left[\mathbf{\Sigma}(f) ^{-1}\right]_{A,A}}=  \hat{J}_{uncorr}^{A,A}(f)=  \sigma_{A}(f)^2 +\left|A_A(f)\right|^2 \sigma_C(f)^2\\
     \textrm{(Intrinsic noise dominated)}
\end{align*}
where we simply recover the naive PSD estimator. Thus, we have derived estimator for the intrinsic (or un-correlated) noise which is equivalent to or better than the naive PSD,
\begin{equation}
     \frac{1}{\left[\mathbf{\Sigma}(f) ^{-1}\right]_{i,i}}=  \hat{J}_{uncorr}^{i,i}(f) \leq \hat{J}_{i}(f).
\end{equation}

Then, one can estimate the correlated noise as 
\begin{align}
    \hat{J}_{corr}(f) = J_i(f) - \frac{1}{\left[\mathbf{\Sigma}(f) ^{-1}\right]_{i,i}}.
\end{align}

To interpret the measured correlated noise calculated using this technique and reported in Section~\ref{sec:discussion_corr_noise}, we investigated the robustness of the decorrelation algorithm in recovering signal information from a noise spectrum that contains a pulse of known energy. We performed a Monte Carlo study that simulates pulse-like events with a fast rise time ($\sim$1~ns, approximating a delta function) and a fall time equivalent to the phonon collection time of the TES for CUTE data ($\sim50$~$\upmu$s). Each pulse was shifted by a time delay that was sampled from a uniform distribution. The pulses were scaled to a range of linearly spaced amplitudes between 0.1 - 10~fW, equivalent to sub-threshold energy depositions for this detector in the CUTE environment. These scaled pulses were injected into time domain power noise traces for both channels which were randomly sampled from the NEP spectrum for CUTE. Events were generated in 16 batches, each of different sample sizes ranging from 100 to 500, adding 25 new samples in each subsequent batch. We then computed the correlated noise for each sample size and pulse amplitude. The reconstructed signal amplitude that was estimated by the decorrelation procedure (A$_{\rm estimate}$) was computed as the mean of the correlated noise spectrum in the frequency range of 10 to 200 Hz where signal-like contributions dominate. 

\begin{figure*}[!h]
     \centering
     \begin{subfigure}{0.48\textwidth}
         \includegraphics[width=1\linewidth]{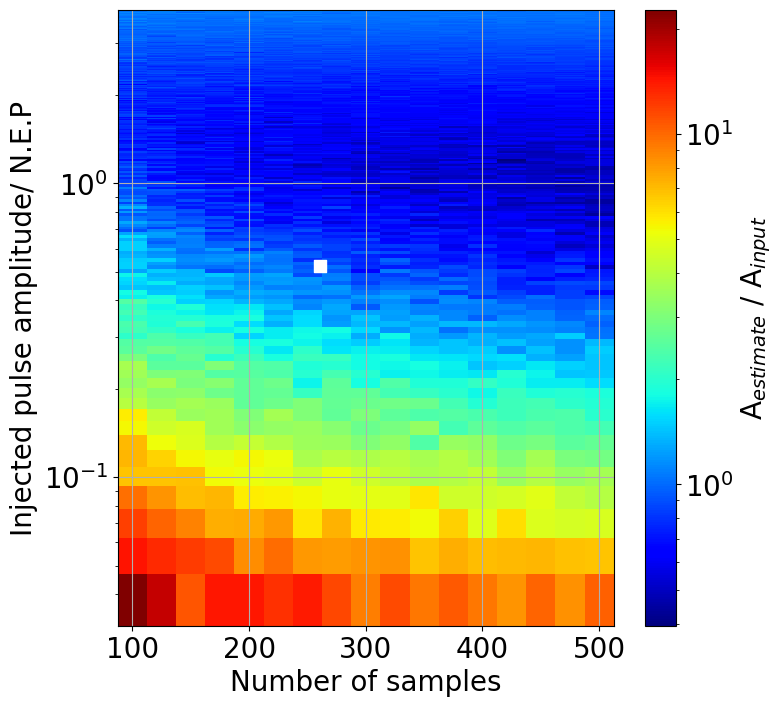}
     \end{subfigure}
     \begin{subfigure}{0.48\textwidth}
         \includegraphics[width=1\linewidth]{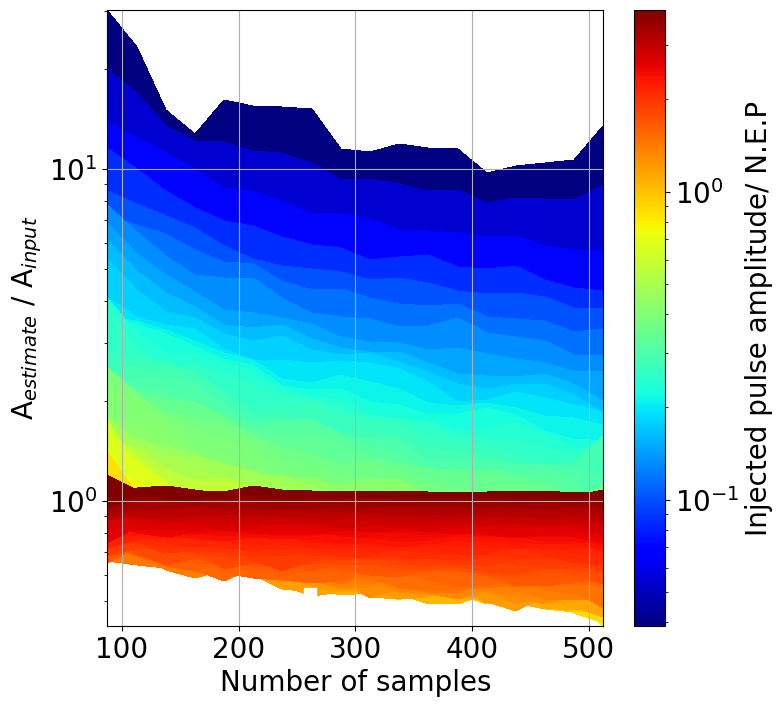}
     \end{subfigure}
     \caption{\textbf{Left.} A heat map showing the distribution of the ratio of the estimated pulse amplitude to the injected amplitude as a function of the number of samples and injected pulse amplitude normalized to the measured NEP. The white square indicates the measured correlated noise from CUTE measured normalized by the NEP measured from 261 samples. \textbf{Right.} This heat map shows the same information as the left plot but with the y- and z-axes swapped. This plot demonstrates that the ratio of the reconstructed to true correlated noise asymptotes to one for large enough sample sizes and pulse amplitudes below the intrinsic noise level. For pulses that are larger than the N.E.P, the algorithm, at worst, seems to underestimate the correlated noise by $\sim$50\% and asymptotes to one for even small sample sizes for injected pulse amplitudes that are $\sim$25\% larger than the noise level.}
     \label{fig:decorr_mc}
 \end{figure*}

Figure~\ref{fig:decorr_mc} shows how the ratio of (A$_{\rm estimate}$) to the mean of the input signal spectrum in the same frequency range (A$_{\rm true}$) is distributed. The heat map is made on a grid of injected signal power normalized to the NEP and the number of samples used to estimate the correlated noise. The plot demonstrates that as injected signal power decreases, a larger number of samples is required to accurately recover the injected signal. For signal amplitudes that are smaller than $\sim$(0.25 - 0.7) $\times$ NEP, the method primarily provides an upper bound for the measured correlated noise, but with increasing sample size and higher injected amplitudes, it begins to yield a true estimate of the signal. For injected power larger than the NEP the technique consistently yields a lower bound. The measured correlated noise from CUTE falls at the boundary of an upperbound or being close to a true estimate. Thus, we establish that our measured correlated noise is at worst an overestimate of the true correlated noise and at best a true estimate and therefore represents a conservative measurement of the true correlated noise.
\end{document}